\documentclass[prl,aps,epsf,showpacs,twocolumn]{revtex4}
\usepackage{times}
\usepackage{graphicx}
\usepackage{float}
\usepackage{latexsym,amsmath,amssymb,bm,euscript}
\usepackage{color}
\usepackage{subfigure}
\usepackage{epstopdf}
\usepackage[colorlinks=true,linkcolor=blue,citecolor=blue]{hyperref}
\usepackage{hyperref}

\begin{document}

\title{Nature of Many-Body Localization and Transition by Density Matrix Renormaliztion Group and Exact Diagonalization Studies}
\author{S. P. Lim and D. N. Sheng} 
\affiliation {Department of Physics and Astronomy, California State University, Northridge, California 91330, USA}

\begin{abstract} 
A many-body localized (MBL) state is a new state of matter emerging  in a disordered  interacting 
system  at  high energy densities through a  disorder driven  dynamic  phase transition.
The nature of the
phase  transition and the evolution of the MBL phase near the transition  are
the focus of intense  theoretical studies with  open issues  in the field. 
We develop an entanglement  density matrix renormalization group (En-DMRG) algorithm to accurately
target the entanglement patterns of  highly excited states for  MBL
systems. By studying the one dimensional Heisenberg spin chain in a random field, 
 we demonstrate the  high accuracy of the method 
in obtaining  statistical results of quantum states in the MBL phase.
Based on large system simulations by  En-DMRG for excited states, we demonstrate some interesting
features in the entanglement entropy  distribution function, 
which    is characterized by two peaks;  one at zero and 
another one  at the quantized entropy  $S=ln2$  with an exponential decay tail on the  $S>ln2$ side.
Combining En-DMRG with exact diagonalization simulations, 
we demonstrate  that the  transition from the MBL phase to the
delocalized ergodic phase is driven by  rare events where   the  locally entangled spin pairs develop long-range power-law correlations.
 The corresponding  phase diagram   contains   an intermediate   regime,
 which has power-law spin-z correlations resulting from  contributions of the rare   events.
We discuss the physical picture for the numerical observations in the intermediate regime, where 
various  distribution functions 
are  distinctly different from  results deep in the  ergodic  and MBL phases for finite-size systems. 
Our results  may  provide new insights for  understanding the  phase transition in such  systems.

\end{abstract}

\maketitle

\section{I.  Introduction}

Understanding the effects of interaction on Anderson localization\cite{basko2006, fleishman1980,altshuler1997,jacquod1997,georgeot1998,gornyi2005}  
 has led to a rapidly expanding
field, where a  new correlated state of matter,  a many-body localized (MBL) phase  emerges\cite{nandkishore2015,altman2015, huse2014,nandkishore2014,oganesyan2007,pal2010,znidaric2008}.
Many remarkable properties of an MBL phase has been established\cite{
nandkishore2015,altman2015, nandkishore2014,oganesyan2007,pal2010,znidaric2008,
rigol2008, serbyn2014,kwasigroch2014,yao2014,vasseur2015,
huse2014, vosk_theory2014, serbyn2013,ros2015,chandran2014,grover2014,agarwal2015, knap2015,
canovi2011,cuevas2012,bauer2013,kjall2014,luca2013,iyer2013,pekker_hilbert2014,johri2014,bardarson2012,andraschko2014,laumann2014,hickey2014,nanduri2014,barlev2014,imbrie2014,groverf2014,ponte2015, huang2015,you2015,serbyn2015,singh2015,barlev2015,deng2015,chen2015,li2015}
 based on  extensive theoretical studies.  
For disordered interacting 
 systems, a  random disorder can drive  a dynamic  phase transition\cite{nandkishore2015,pal2010,potter2015trans} from 
a delocalized state to an MBL phase,  where  all  energy eigenstates  become   localized.
Protected by the localization, an MBL phase is non-ergodic and can not thermalize\cite
{deutsch1991,srednicki1994,rigol2008}, which also challenges the fundamental ``eigenstate  thermalization hypothesis'' (ETH)  for quantum statistical  physics\cite{hosur2015}. 
The  energy  eigenstate in an MBL phase has
 entanglement entropy satisfying an area law\cite{nandkishore2015, bauer2013,kjall2014,  grover2014}
scaling  in contrast to the volume law scaling expected for an ergodic   delocalized state.  
The MBL phase behaves like  integrable systems, respecting   extensive numbers  of local conservation laws 
~\cite{serbyn2013,huse2014, ros2015,chandran2015} with the emergence of the localized-bits (l-bit) representing these
conserved local degrees of freedom.  Interestingly, 
exotic topological states    usually present   at low temperature,
can survive  to infinite temperature in an MBL environment\cite{huse2013,bahri2013,chandran2014,bauer2013,vosk2014,pekker_hilbert2014, potter2015,yao2015}, which greatly enhances the 
possibility of their applications  in future topological quantum computing.
There are also growing experimental activities probing  
the nature of the  MBL phase and phase transition in cold atom systems~\cite{serbyn2014,kwasigroch2014,yao2014,vasseur2015, bordia2015}.

So far,   theoretical understanding of the dynamic  phase transition is still
at the beginning stage\cite{oganesyan2007,pal2010,znidaric2008,canovi2011,cuevas2012,bauer2013,serbyn2014,kjall2014,vosk_theory2014,luca2013,iyer2013,pekker2014,johri2014,bardarson2012,andraschko2014,laumann2014,vasseur2015,hickey2014,nanduri2014,barlev2014,grover2014, luitz2015, serbyn2015,goold2015}. 
Larger sizes (with up to $N=22$ spins) numerical  exact diagonalization (ED) studies\cite{luitz2015}  of
the 1D Heisenberg chain in a random field have  demonstrated a continuous  phase transition 
between a delocalized ergodic  phase  to an MBL  phase based on extensive finite-size scaling analysis
of different physical quantities including the  entanglement entropy  and  the energy level statistics.
The numerical linked cluster expansion calculations suggest a higher critical disorder strength  for entering the MBL phase\cite{devakul2015} 
  than that obtained by ED studies\cite{luitz2015}. 
Theoretical\cite{vosk_theory2014}  and numerical  studies of the  low frequency conductivity\cite{agarwal2015, knap2015} and 
energy spectra statistics\cite{serbyn2015} have  
suggested that there is an intermediate regime  with sub-diffusive conductivity
and (or)  semi-Poisson level statistics  between the ergodic and MBL phases.
A consistent picture for understanding the dynamic  phase transition in such a system is still absent.
One of the  difficulties  is the presence of  rare Griffiths regions\cite{vosk_theory2014, potter2015trans, knap2015}
which may have  singular contributions in  driving a  phase transition. However, so far there is
still limited  quantitative understanding about  their effects.

To make progress, it is highly desirable to  study much larger systems\cite{chandran2015finite}
and to  establish  the nature of the MBL phase in the thermodynamic limit,
which are  great challenges for such a correlated system at finite energy density.
The MBL phase has low entanglement similar to  groundstates of low dimensional systems,  which has stimulated a lot of recent effort in developing the density matrix renormalization group (DMRG)\cite{white1992}
or tensor network based new algorithms for studying such systems\cite{pekker2014_mps,friesdorf2015,pollmann2015,chandran2015_construct,khemani2015,yu2015}.
Exciting progress has been made including  developing modified DMRG methods to accurately target  some of the highly excited eigenstates demonstrated by two  recent 
preprints\cite{khemani2015,yu2015}. 
 One of the main issues that remains to be addressed  is if it is possible to  use  the DMRG method to unbiasedly obtain 
different excited states with intrinsically fluctuating entanglement entropy for large systems. 
Only when the DMRG method for excited states can overcome the tendency of picking minimum entangled states\cite{jiang2012}  among all excited
states at a finite energy density, will  it  establish  itself as  
 a powerful tool for studying challenging and fundamental issues in   quantum statistics 
emerging with the MBL phase.

In this article, we report  developing  a new  entanglement DMRG (En-DMRG) algorithm to meet this challenge.
The En-DMRG will randomly select and  target the entanglement pattern of the highly excited states in an MBL system. 
By studying the one dimensional Heisenberg spin chain in a random field,  we demonstrate the  high accuracy  of the method 
in reproducing  statistical features of the  system in  comparison with ED results.
Based on large system simulations with up to $N=72$ spins by En-DMRG, we first show that a
spin-flip process and the associated spin-entangled pairs   have a finite  and system-size independent probability density in the MBL phase. 
 We also obtain the characteristic probability density distribution function  
for the entaglement entropy,  which has   a continuous   spectrum with a sharp peak at the quantized 
entropy value $S=ln2$  and  an exponential decay tail on the $S>ln2$  side for the MBL phase.
 Combining En-DMRG with ED simulations, we study the driving force of the dynamic phase transition from the MBL phase
to the ergodic phase.  We find that the  spin-entangled pairs 
first become long-range power-law entangled, which leads to a strong enhancement in the probability distribution function of
the entanglement entropy (fluctuation of half system magnetization) on the larger entropy (fluctuation) side.
We also identify an intermediate regime where   the rare events contribute significantly to the average of
the spin  correlations.
Our results  may provide new insights  for understanding the rich physics of 
the MBL phase and the exotic dynamic phase transition in such  systems.

\section{II.  En-DMRG Method for Highly Excited States}

The standard DMRG\cite{white1992} created by White is an unbiased and controlled method  for obtaining ground state
or a few low energy excited states of interacting systems.  The true power of the method
is in  its way of constructing  the HS by using the eigenstates of the
reduced density matrix. To target  excited states in an MBL system,
we develop an  En-DMRG method  based on the standard DMRG  with modified
initial  sweeping process to optimally construct larger HS for these  states.
During this process, we use a varying  bond dimension to allow a natural development of an entanglement structure
for the quantum state, which leads to  a rapid convergence of the entanglement entropy.
Here, we outline the basic steps of the  En-DMRG.  (i) We start from the  standard DMRG\cite{white1992}
using the  ``infinite'' process. 
(ii) Once our system reaches the required system size with $N$ spins, we start the sweeping process to build the
HS  with a varying bond dimension. 
(iii) We use the Lanczos method to obtain
the lowest  few energy eigenstates of the squared Hamiltonian $(H-E_t)^2$ ($E_t$ is the target energy) with lower accuracy
and keep up to three eigenstates  for the reduced density matrix\cite{white1992}
in the initial  few sweeps to build the HS
 until the wavefunction starts to vary smoothly with sweeping. 
 (iv) After building the  HS,   we perform the standard sweeping with much higher accuracy for targeting the lowest
energy eigenstate of the squared Hamiltonian  until the targeted state   is converged. 
One may also use other eigenstate solvers for optimization\cite{khemani2015,yu2015}.

Conceptually important for our  En-DMRG method, 
we use  a varying and larger  bond dimension to stimulate the expansion of the entanglement in the initial ``infinite'' and the beginning few  DMRG
sweeps, which builds the HS for   different eigenstates with a  range  of entanglements.
We start from targeting one state near $E_t$.  We increase the bond dimension and target  more
states for the reduced density matrix when the sweeping gets frustrated (several new wavefunctions have large
overlaps with the initial wavefunction at the previous step).
Once we have built a proper HS for the excited state with its entanglement pattern established, only a few steps of   Lanczos iterations 
is required to  find an accurate eigenstate using the wavefunction transformation following the standard DMRG\cite{white1992}.
 The aim  of our work  is to establish the high accuracy of 
the En-DMRG for quantum statistics for {\it large} MBL  systems 
as we will demonstrate  below, which has not been addressed\cite{khemani2015,yu2015}.
In combining with the ED simulation on the smaller
disorder side (where the DMRG method is not applicable because of the volume entanglement and
the similarity for nearby eigenstates) for the delocalized phase, we will explore the nature of the MBL phase and its transition.
\\

\begin{figure}[b]
 \includegraphics[angle=-90,width=4.95in]{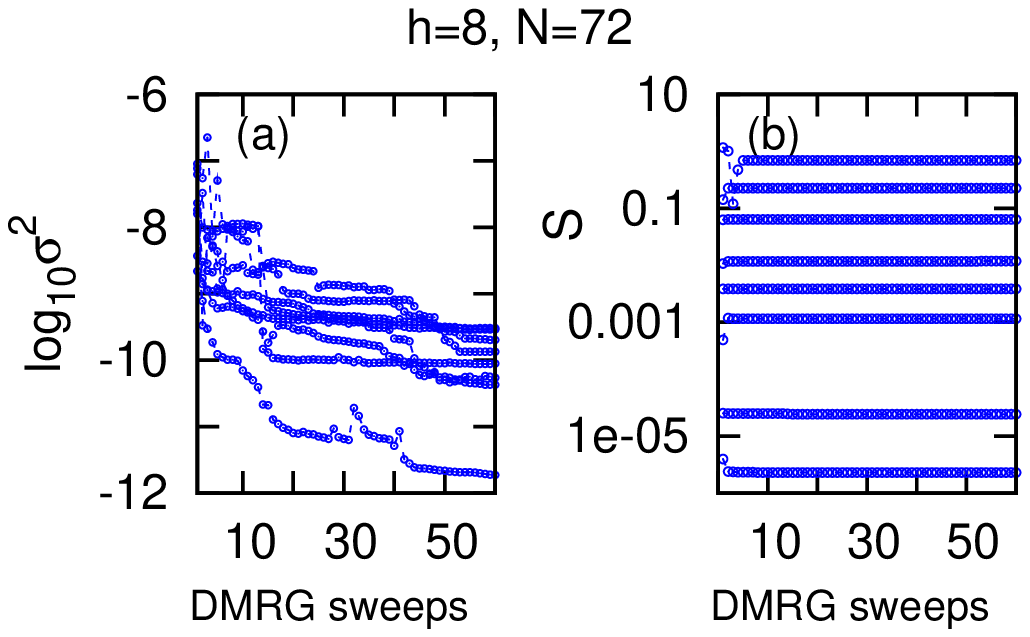}\\
 \includegraphics[angle=-90, origin=c, width=3.7in]{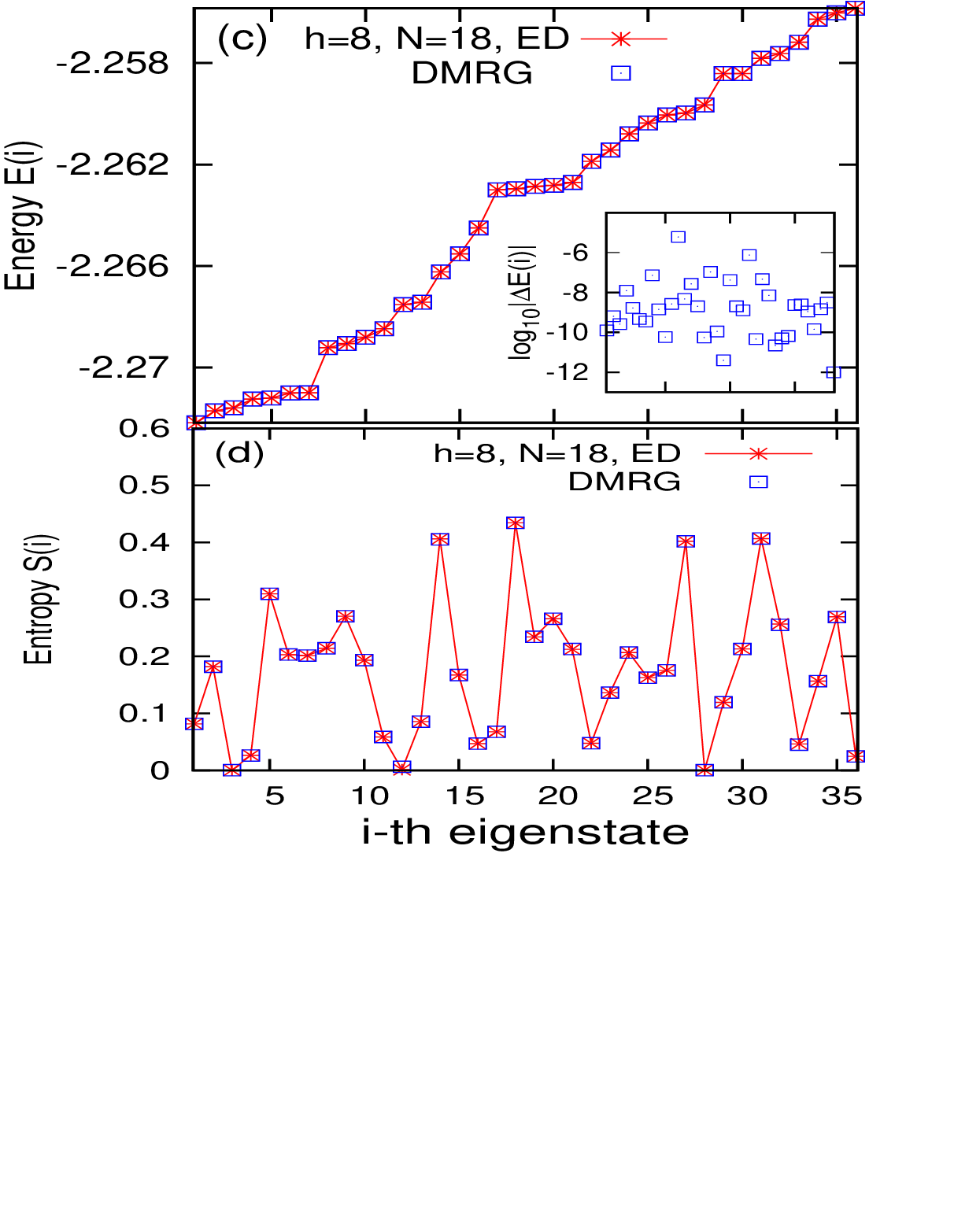}\\
\vspace{-1.3in}
\caption{(Color online) (a) The logarithm of the energy variance $\sigma^2$ 
as a function of the number of En-DMRG sweeps for 8 randomly selected
En-DMRG processes for system size  $N=72$ and $h=8$ targeting  the
state at the middle of the energy spectrum. 
(b) The entropy $S$ evolution for the same processes. 
(c) For smaller system $N=18$ at $h=8$,  we show the energy eigen values
for $i-th$ eigenstate
found in the energy  interval $(-2.272, -2.257)$ near the energy spectrum center,
where an excellent  agreement between the En-DMRG and ED results is  found.
The absolute energy difference $|\Delta E|$  between En-DMRG and ED eigenstates are shown in the inset of
(c), which typically  is around $10^{-9}$.  
(d) For all 36 states we found in (c),   we demonstrate their one to one correspondence for
entanglement entropy $S$ between ED and En-DMRG results.
 }
\label{fig1}
\end{figure}

We  study  the Heisenberg  spin-1/2 chain   
with  the following Hamiltonian:
\begin{equation}
 H =  \sum_{i=1}^{N-1} \vec{S}_i \cdot \vec{S}_{i+1}
  -  \sum_{i} h_iS_i^z,\nonumber
\end{equation}
where the nearest neighbor 
 coupling  $J=1$  sets   the 
energy scale  and we use open boundary for a  better convergence  in DMRG.
The $h_i's$ are the random magnetic field couplings, which  distribute  uniformly  in
the interval $(-h,h)$ with $h$ as the strength of  random fields.
The ED studies using system sizes  $N=12-22$ established an MBL phase at $h \gtrsim 3.5$\cite{luitz2015}.
The En-DMRG allows us to study larger  systems   up to $N=72$ spins, which makes the
extrapolation to the thermodynamic limit for the MBL phase possible.
All the results are obtained near  the center of the energy spectrum.

 To demonstrate  the convergence of the process for our larger system calculation with $N=72$ at $h=8$,  we show the evolution of the logarithm of the energy variances 
$\sigma^2=\langle H^2 \rangle-\langle H \rangle^2$ in Fig. 1a  for each sweep of the En-DMRG process, which usually is proportional to the
energy error of the  state obtained by the En-DMRG.
Eight different En-DMRG targeting runs are illustrated in Fig. 1  from one random disorder configuration,
and we use an initial bond dimension $M=24$ and vary it up to  $M=64$ during the sweeping process.
We find that the $\sigma^2$ starts from around $10^{-8}$ for the first a few sweeps and drops to around 
$10^{-10}$ within  60 sweeps, where
we complete the En-DMRG process to find an energy eigenstate near the target energy $E_t=-1.1$.
The corresponding bipartite
entanglement entropies  are shown in Fig. 1b, where amazingly we find
that entropy $S$ is well converged after a few initial sweeps.  As also demonstrated here, the eigenstates
we obtain have a reasonable  variance of $S$ values reflecting different levels of entanglement of the targeted
eigenstates. 
To benchmark our results  at a smaller size $N=18$ and $h=8$,
we show the comparison between eigenenergies obtained  using ED (red star) and
En-DMRG (blue box) with a varying bond  dimension around  $M=16\sim 48$,  
where an excellent  agreement is demonstrated
for all energy eigenstates in the energy interval  $(-2.272, -2.232)$ near the energy spectrum center 
for one disorder configuration.
The absolute energy difference $|\Delta E|$  between En-DMRG and ED eigenstates are shown in the inset of the
Fig. 1c, which typically  is around $10^{-9}$.  
In En-DMRG targeting we run about
1000 times with slightly different targeting energy  $E_t$ each time. Some states are found  more frequently than other states.
In Fig. 1d,    we show  their one to one correspondence for
entanglement entropy  $S$ from ED and En-DMRG calculations.
Very interestingly, we also find that entropies intrinsically fluctuate for these states with very close energies,
and En-DMRG can capture them all precisely.   However, due to the unequal appearance of the different
eigenstates in En-DMRG runs,  one needs to address the ability  for En-DMRG to capture the statistics of the system for different
physical quantities in comparison with  ED results\cite{luitz2015}.
We will demonstrate the success of En-DMRG in this aspect as we present new results  below.   

\begin{figure}[b]
 \includegraphics[angle=-90,width=3.9in]{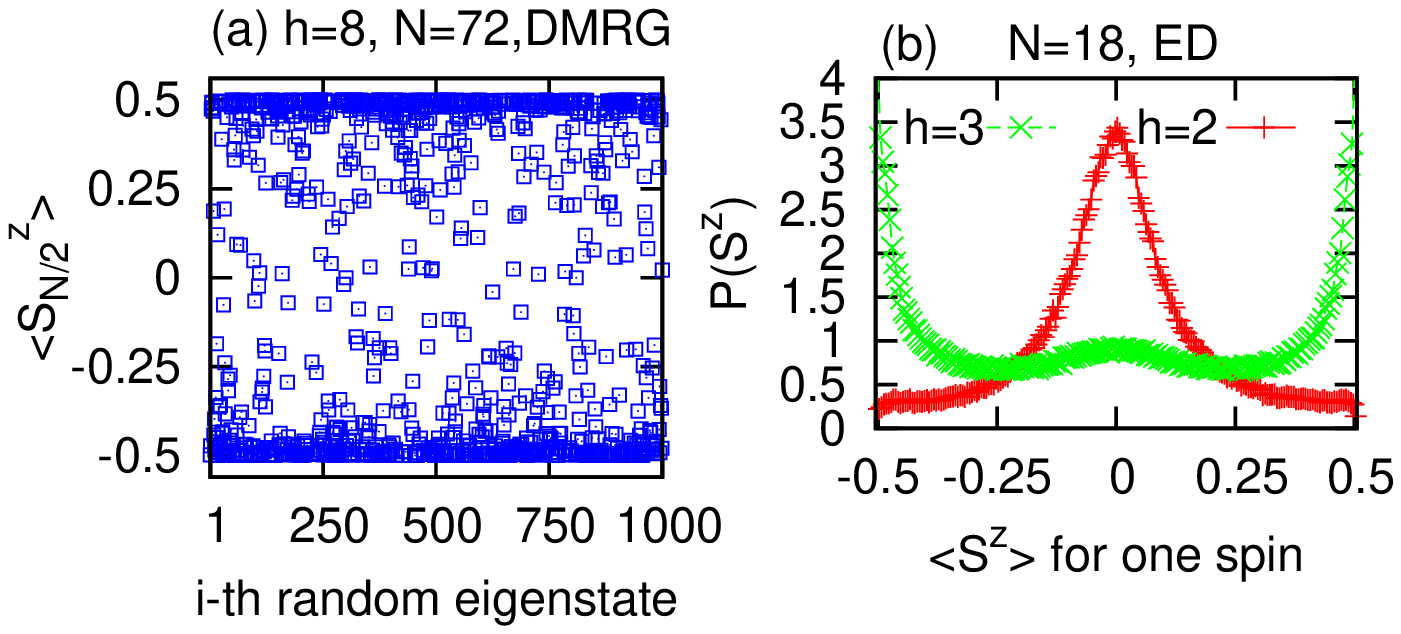}
 \includegraphics[angle=-90,width=3.9in]{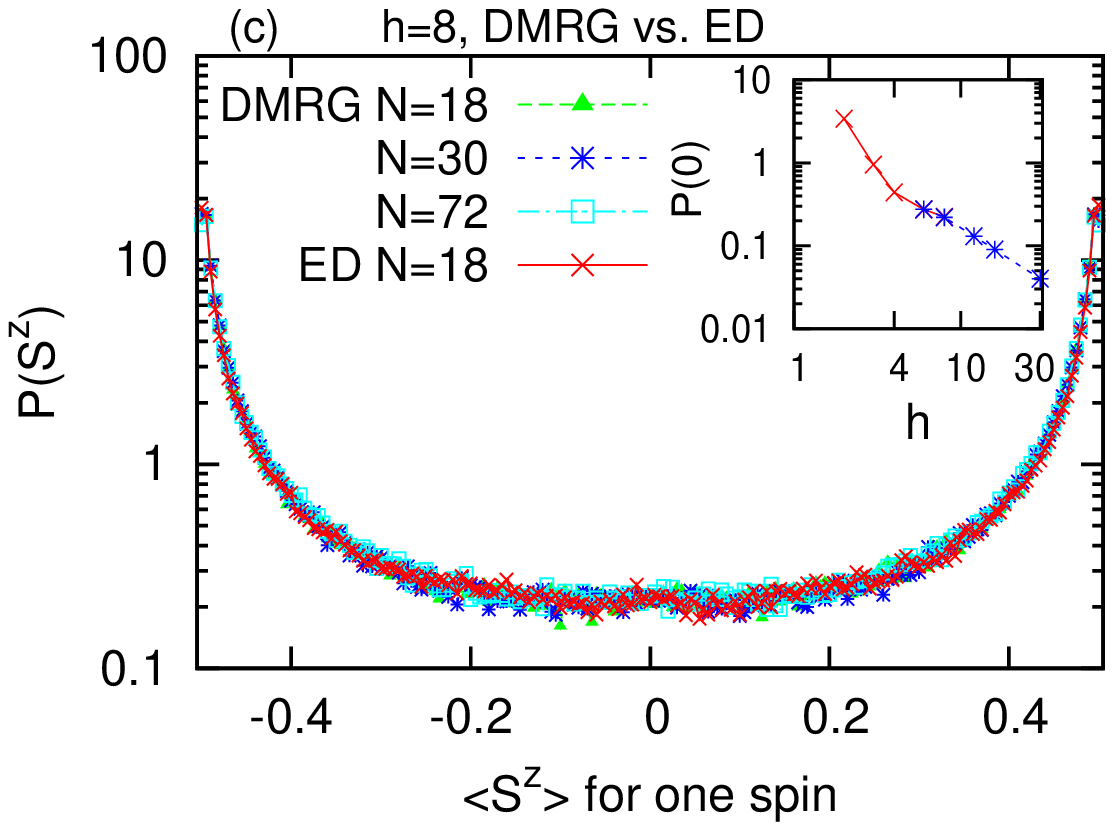}\\
\caption{(Color online) (a)  
The spin-z expectation values $\langle S_{N/2}^z\rangle $ for $N/2$-th site 
for eigenstates obtained by  En-DMRG  
 for 1000 random disorder configurations at $h=8$ and  $N=72$.
(b) The probability density distributions $P(S^z)$ for  systems with weaker disorder $h=2$ and $3$
obtained by ED at $N=18$  using an ensemble of 2000 disorder configurations with 50 states 
from each configuration.
$P(S^z)$ has  the Gaussian distribution for $h=2$ 
while there are two peaks at $\langle S^z \rangle \sim \pm  0.5$ for $h=3$.
(c) $P(S^z)$ for different system sizes from  $N=18$ to $N=72$ for strong disorder
$h=8$ using an ensemble of 20000 disorder configurations.  
 The sharp peaks at $\langle S^z \rangle \sim \pm 0.5$ demonstrate the violation of the ETH  for this system. 
The results obtained by  En-DMRG  and ED are near identical at $N=18$ establishing the unbiased sampling
 for our En-DMRG. 
Shown in the inset is $P(0)$  as a function of $h$.
For (b-c), the typical standard error bar (obtained by dividing ensembles into 10 groups) is about the size of symbols.
 }
\label{fig1}
\end{figure}

\section{III. Numerical Results}

{\bf A. The spin polarization and spin-flip}

One of the characteristic features of the MBL phase is that there is
a set of localized l-bits, which represents
$N$ locally conserved and commuting effective  spins\cite{huse2014, serbyn2013,vosk_theory2014,ros2015,chandran2015}.
In general, these l-bits are locally dressed versions of the underlying physical  degrees
of freedom. 
To address  the microscopic nature of l-bits and their evolution near the phase transition, we  first    examine the
probability density distribution function of each spin for large systems.  
As shown in Fig. 2a,  we plot  
the spin-z expectation value $\langle S_{N/2}^z \rangle = \langle \Psi|S_{N/2}^z|\Psi \rangle $ for the middle ($N/2$-th) site 
for the En-DMRG obtained  eigenstate $|\Psi \rangle$ with total $S_{tot}^z=0$   for  1000 random disorder configurations  
at   $N=72$ and $h=8$ (we take one state from each configuration).
It shows a wide distribution with much enhanced appearance near $ \langle S^z \rangle=\pm 0.5$ quite different from an ergodic  state where the single site distribution should 
take on  values close to the averaged value  which is zero here. 
We have checked that spins at all other sites  show very similar pattern.     For comparison, 
we first show the  probability density distribution $P(S^z)$ in Fig. 2b for  systems with weak disorder $h=2$ 
and $3$ 
obtained by ED at $N=18$ averaged over 2000 disorder configurations with 50 states near energy spectrum center 
from each configuration\cite{luitz2015}.
We find that $P(S^z)$ has the Gaussian distribution  in the ergodic phase
at $h=2$ with a strong peak around the value $ \langle S^z \rangle =0$, which also grows sharper approaching
a delta function with the increase of $N$.   The distribution at
$h=3$ is quite different with a very  broad structure and peaks near $ \langle S^z \rangle \sim \pm 0.5$, which also shows
a very weak $N$ dependence.  Now we obtain the distribution function $P(S^z)$ for systems with different sizes $N=18$, $30$, and $72$
in the MBL phase with $h=8$ using 20000 disorder configurations and one state from  each configuration 
obtained by  En-DMRG as shown in Fig. 2c.
We see a completely different distribution compared  to the $h=2$ result (but similar
to $h=3$ case qualitatively), with sharp peaks at $\langle S^z \rangle =\pm 0.5$ demonstrating
the violation of ETH  for this MBL system. The results are  system size independent and fully converged 
  suggesting the same distribution in the thermodynamic limit.
By comparing the  results obtained by  En-DMRG   and ED  at  $N=18$ for $h=8$  (in Fig. 2c) we also establish
that our En-DMRG is  unbiased, which reproduces all the different events 
 with the right probability as they appear in the random quantum systems. 
 In the $P(S^z)$ distribution function, we also see the finite probability density
 $P(0)\sim 0.21$ for   $ \langle S^z \rangle =0$. 
We can understand these regions in a perturbative way.  The perturbation from the xy-coupling  in the Hamiltonian 
 gives rise to  
   spin-entangled pairs in the near polarized spin background, which we refer to as  spin-flip events. 
The  nonzero probability  for   $ \langle S^z \rangle =0$  is a consequence of  these spin-entangled pairs. 
We show $P(0)$  as a function of $h$ in the inset of Fig. 2c, which is generally nonzero for finite  $h$ approximately
following $P(0)\sim 1/h$ on larger $h$ side (we obtain results from ED in the
smaller $h$  side)
revealing   the significance of  spin-flip events in the  whole MBL phase.
Because the  probability density $P(0)$ is system size independent,  then at large $N$ limit,  there will be a finite
density for the spin-entangled pairs.

{\bf B. Distributions of entanglement entropy  and fluctuation of half system magnetization}

The bipartite entanglement entropy $S$ has been extensively used as an effective tool  to
characterize  different many-body states  for such an  interacting system\cite{nandkishore2015, luitz2015, kjall2014}.
 We compute the Von Neumann entanglement entropy   from all eigenvalues of the reduced density matrix
$\rho_A$ as
 $S=- {\rm Tr} \rho_A \ln \rho_A$,
 by partitioning the system in  the middle of the spin chain.  Different from the general 
 volume law entanglement entropy for  ergodic phase on the weak disorder side,
  the MBL phase at $h \gtrsim 3.5$ side 
has  the entanglement entropy following the area law, which is the fundamental reason that  DMRG can work for such a phase.
We now   study  the  probability density distribution  of the entropy $P(S)$
for spin system near the energy spectrum center. 
We choose the statistical ensemble from at least $50000$ disorder configurations for DMRG calculations for each $N$.  
As shown in the Fig. 3a,  results from system sizes with $N=18$, $30$ and $72$ are on top of each other, 
suggesting the same distribution in the thermodynamic limit.
The agreement between the En-DMRG  and ED  results for  $N=18$ systems
 confirms  the robustness
of En-DMRG in capturing all different states with a wide range of the entropy distribution.
 The $P(S)$ results  in the whole $S$ region are converged (independent of the bond dimension) except for a couple of  data points 
near $S\sim 0$, where
the DMRG results  are  a few percent smaller than the ED results for the larger bond dimensions we used. 
While this small difference does not change the universal behavior of the $N$-independent distribution,
 the error  is caused by the fact that these states are close to product states (which are slightly harder to be captured by En-DMRG).
Focusing  on the characteristic feature  of  $P(S)$,
we find that  $P(S)$ is peaked at $S=0$ with a continuous spectrum going into the finite $S$ range,
and with a second peak at a quantized value $S=0.69 \sim ln2$.   Quantitatively, the $P(S)$  value at the second peak
is about $1/20$ of $P(S=0)$, which is only clear  in our logarithmic plot.
Quite interestingly, the $P(S)$ has a smooth plateau feature on the $S< ln2$ side, but shows a large  
exponential drop on the $S>ln2$ side. Similar distributions are obtained for larger $h$, where $P(S)$ values goes to zero on the $S>ln2$ side exponentially fast
with the increase of $h$. 
Comparing to the observation of the spin-flip process in the $S^z$ distribution,
we can attribute the $ln2$ peak to these small density  spin-entangled pairs crossing  the two half systems.
The continuous spectrum comes from the
local interaction of spin-entangled  pairs  with surrounding polarized spins,  which  partially reduces their 
entanglement as they become dressed.  The exponential drop of $P(S)$  on the $S>ln2$ side suggests that 
the events of  different  spin-entangled pairs getting entangled together or multi-spin resonant  states 
   have  exponentially small probability 
for  $h=8$ deep in the MBL phase.  
 Our method is also distinctly different from other modified DMRG methods for excited states\cite{khemani2015} as it can accurately describe
these rare regions  with entangled spins  in the center.

\begin{figure}[b]
 \includegraphics[angle=-90,width=3.6in]{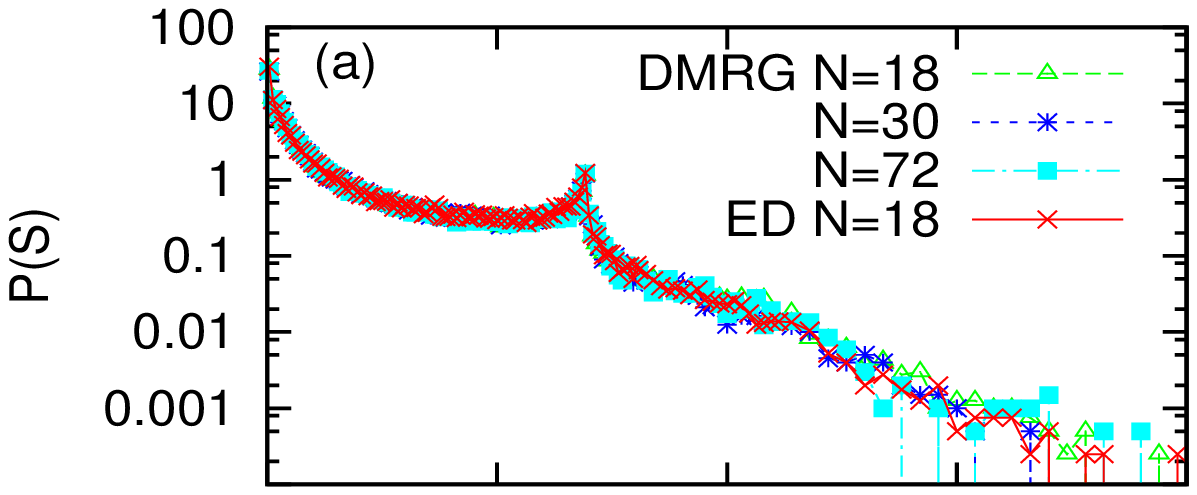}\\
\vspace{-0.20in}
 \includegraphics[angle=-90,width=3.75in]{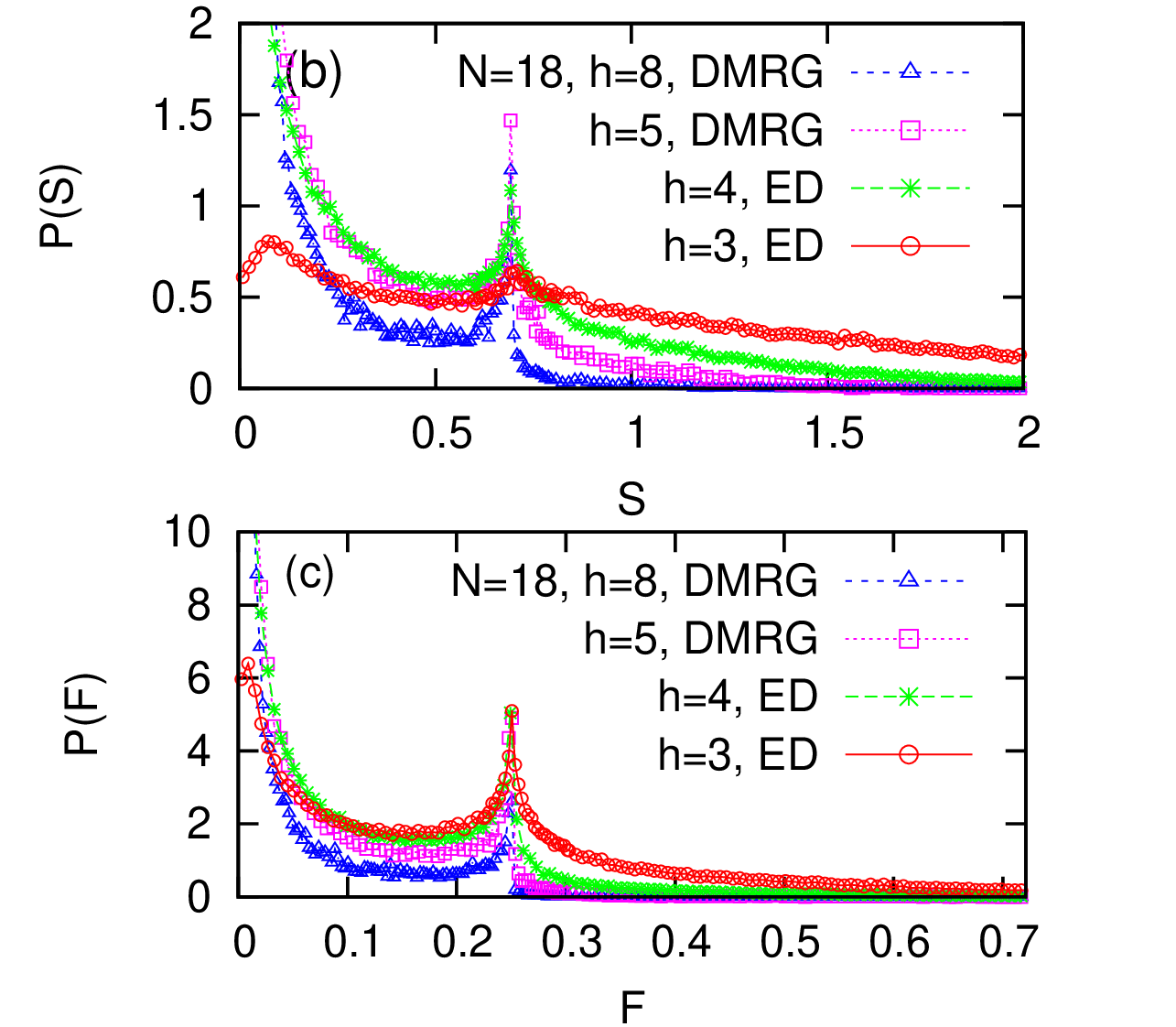}\\
\hspace{0.0in}
\vspace{0.20in}
\caption{(Color online) (a) The entropy probability density distribution 
function $P(S)$  for  $N=18$, $30$ and $72$ at $h=8$. 
For  $N=18$, ED and En-DMRG results find excellent agreement.
(b) The evolution of  $P(S)$ from the MBL phase to the delocalized phase
by reducing $h$ from $8$ to $3$ for  $N=18$ from both ED and En-DMRG calculations. 
All curves have a peak
at or near $S=0$ and a second peak near $S=0.69\sim ln2$. 
(c) The evolution of the probability density $P(F)$ for the variance of the half system 
magnetization $F$ for $h=8, 5, 4$, and $3$.  The second peak is located at a quantized 
value $F=1/4$. 
For results shown in Fig. 3, we use at least $50000$ energy eigenstates for each system size.
The typical standard error bar for (a-c) is about the size of symbols and it is larger near the tail of  distributions.
}
\label{fig1}
\end{figure}

Now we show  the evolution of the $P(S)$ into the ergodic phase 
by reducing $h$. 
The distribution $P(S)$ for $h=8$ to $4$ are  qualitatively similar as shown in Fig. 3b, 
 with  a peak at $S=0$ and a second peak near $S=0.69\sim ln2$. 
However $P(S)$ shows  much enhanced
weight at $S>ln2$ side with reducing $h$ (results at $h=3$ and $4$ are obtained by ED). 
At $h=3$,   the $P(S)$ shows some   different features with first peak
moving away from zero to $S\sim 0.1$ and a long tail
at $S>ln2$ side, which  only decays very slowly.  
We can still see a weak  peak at $S=ln2$, which is very broad and about to disappear. 
Interestingly, these features for $P(S)$ at $h=3$ is still quite different\cite{serbyn2015}  from the Gaussian form
 with a single peak, which is the case
for  $h=2$  with a sharp peak at  the value $S \sim 4.7$  as we checked.

To further establish the above  picture,   we investigate the probability density distribution
$P(F)$ of the fluctuations  of the magnetization of the  half system\cite{luitz2015} 
defined as  $F=\langle (S_h^z)^2 \rangle - \langle S_h^z \rangle ^2$ calculated using the  eigenstate,
where $S_h^z$ is the total spin-z component of the half system. 
If the half-system cutting through a spin-entangled pair  while all other spins are short-range correlated (or near polarized),  then we expect
the variance $F=1/4$. Interestingly, we see that the distribution $P(F)$ indeed has a second peak 
at the quantized value $F=1/4$, which can be attributed to the spin-entangled pairs (local spin flips).  
The overall structure of $P(F)$ is very similar to $P(S)$ with the broad continuum at $F<1/4$ side and a tail into larger $F$
side with its magnitude growing with the reducing of $h$.   
At $h=3$, the second peak is more robust in $P(F)$ than  in $P(S)$ indicating the faster growing of the
entanglement near the phase transition for finite-size  systems\cite{singh2015}.

{\bf  C. Spin spin correlation function and many-body phase diagram}

\begin{figure}[t]
\vspace{10mm}
 \includegraphics[angle=-90, width=1.02\linewidth]{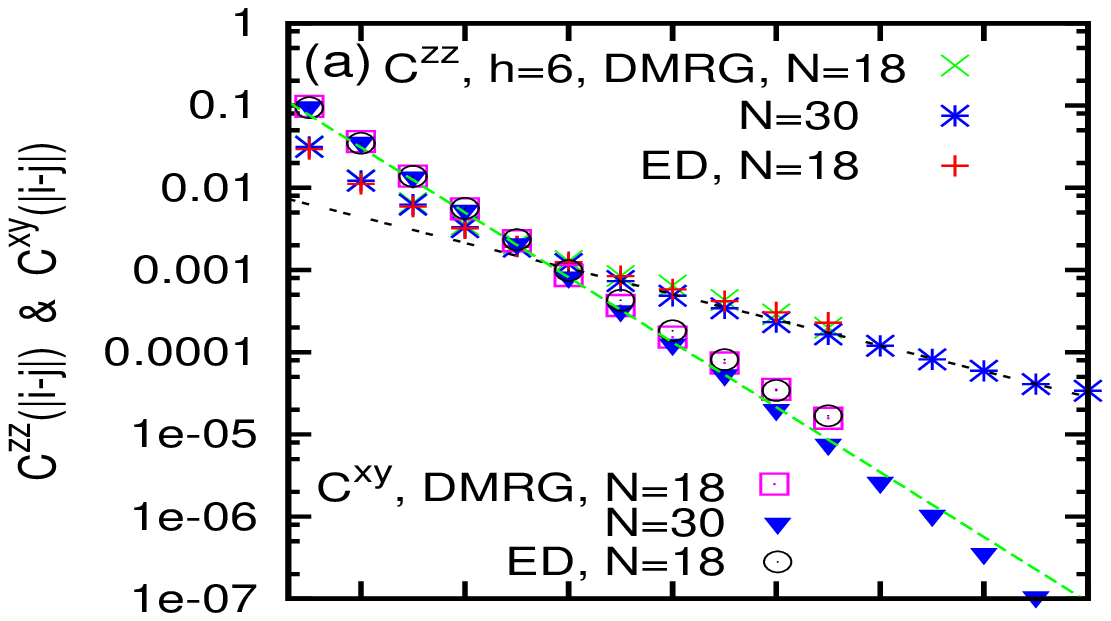}\\
\vspace{-8mm}
 \includegraphics[angle=-90,width=1.02\linewidth]{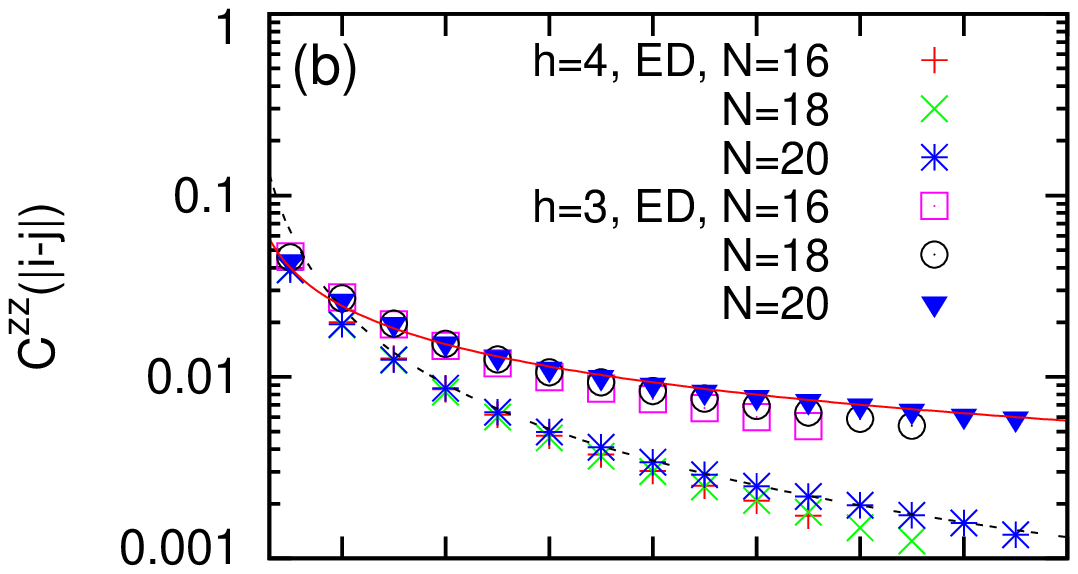}
 \includegraphics[angle=-90,width=1.02\linewidth]{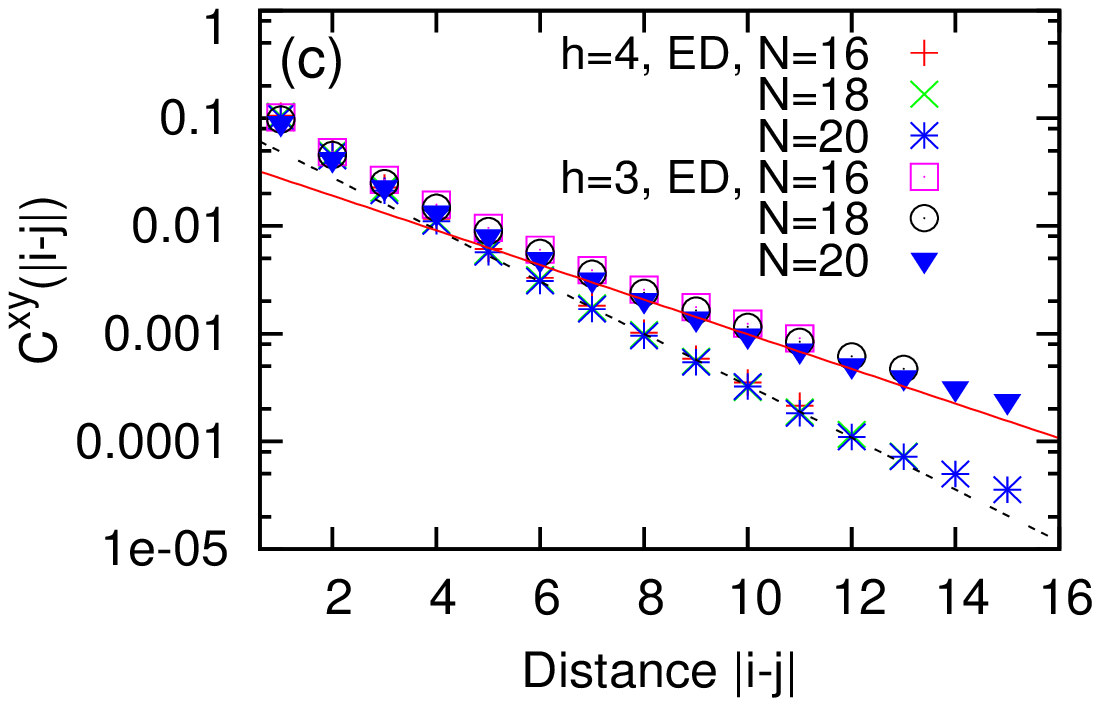} \\
\vspace{-30mm}
 \includegraphics[width=0.93\linewidth]{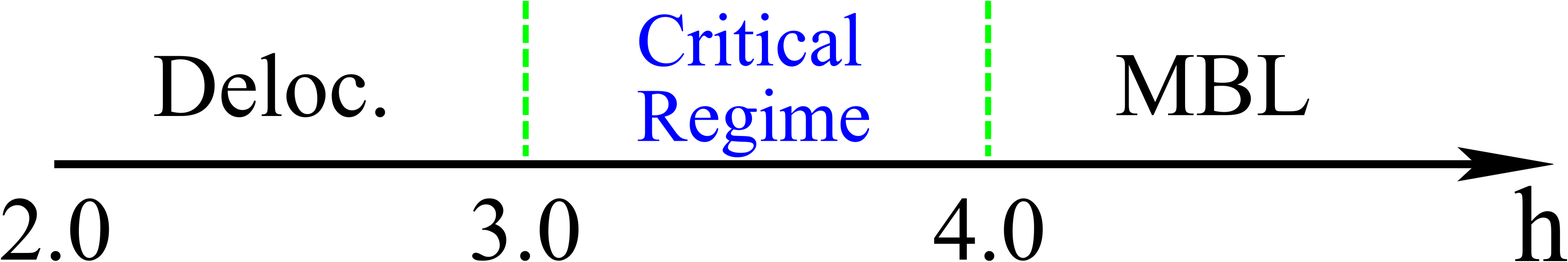}\\
\caption{(Color online) (a) 
The disorder and real space averaged spin-z   $C^{zz}(|i-j|)$ and spin-xy $C^{xy}(|i-j|)$
correlations 
obtained from  En-DMRG at $h=6$ for  $N=18$ and  $N=30$.
(b) The  $C^{zz}(|i-j|)$ develops power-law correlations for $h=3$ and $4$ obtained by
ED.
(c) The hopping correlations $C^{xy}(|i-j|)$ decay  exponentially at $h=3$ and $4$ obtained
by ED. 
(d) A phase diagram with  a delocalized ergodic phase, an intermediate  
critical Griffiths regime, and an MBL phase. See more discussions on the
 finite-size effect of the critical regime in the main text.
\label{fig2}\\
}
\vspace{-0.30in}
\end{figure} 

From the entropy distribution function we have  seen that 
 multiple spins   getting entangled with each other  have exponential small probability deep inside an MBL phase,  which grow
 with reducing $h$ toward the transition region. 
Here, we seek a better understanding of this feature by calculating the disorder averaged spin-z correlations\cite{pal2010, motrunich2000} 
$C^{zz}(|i-j|)=\overline {|\langle \Psi|S_i^zS_j^z|\Psi \rangle - \langle \Psi|S_i^z|\Psi \rangle \langle \Psi|S_j^z|\Psi \rangle|}$ and
spin transfer  (transverse correlations)
$C^{xy}(|i-j|)=\overline {| \langle \Psi|S_i^+S_j^-|\Psi \rangle|}$ ($\Psi$ is the excited eigenstate and the over-line represents the
disorder and real space average).     
We find  that  typical spin correlations are decaying  fast even for intermediate disorder strength $h\sim 4$. However,  
there are rare configurations where the spin correlations at larger distance are strongly enhanced, which 
may be related to  rare Griffiths events\cite{vosk_theory2014, potter2015trans, knap2015, motrunich2000}.  
The arithmetic  average we use here allows the rare Griffiths events to have singular contribution to the correlations
near the transition region.
First, we show the exponential decay behavior for these correlations
in the MBL phase at $h=6$
in Fig. 4a for  $N=18$ and $30$.  We find that the $C^{xy}(|i-j|)$ decays much
faster than $C^{zz}(|i-j|)$ with  
 correlation length $\xi_{xy}=1.1$ and $\xi_z=2.78$, respectively.
Again, the ED results at $N=18$  agree very well  with the En-DMRG results.

Now we move towards  the transition region by reducing $h$ and performing ED calculations.   Shown in Fig. 4b for 
spin-z  correlations,  $C^{zz}(|i-j|)$ is best fit by a power-law 
function  $C^{zz}((i-j|)\propto 1/|i-j|^{\alpha}$ with the correlation exponents  $\alpha=0.7$ and
$1.4$ for $h=3$ and $4$, respectively. 
 The fitting is more robust for larger $N=20$ data.
In Fig. 4c,  we demonstrate the spin transverse correlations for these  systems, where a  clear exponential
decay $C^{xy} \propto exp(-|i-j|/\xi_{xy})$  is  observed   with correlation
 length $\xi_{xy}=2.7$ and $1.8$ for $h=3$ and $4$, respectively.
These results clearly establish that there is an intermediate regime 
around $h\sim 3-4$, which  has  exponential decay spin transfer demonstrated  from $C^{xy}(|i-j|)$, while
the entanglement grows through  correlations of  different  spin-entangled pairs 
seen in the power-law $C^{zz}(|i-j|)$  correlations. 
This is also consistent with the entropy  distribution function, where
the strong power-law  tail develops on the   $S>ln2$ side for
these intermediate $h$. 
Based on these results, we obtain  a phase diagram shown in Fig. 4d,  where we find 
delocalized ergodic phase,  a critical Griffiths regime,  and an MBL phase with increasing
$h$.  From finite size results,  we estimate the  critical regime   is between  
$h\sim 3.0-4.0$
 based on the exponential decaying behaviors of $C^{xy}$ before entering
ergodic phase at smaller $h$ side, and also the power-law behavior of the $C^{zz}$ entering the critical
Griffiths regime.   

\section{IV. Summary and discussion}

Based on the newly developed En-DMRG method for excited states of MBL systems, we establish  the thermodynamic distribution functions for spins,
 entanglement entropy and  fluctuations of the half system magnetization, 
and demonstrate  the physical picture of the MBL state.
 We study  the  dynamic phase transition from the MBL phase to the ergodic delocalized phase
and find  there is an intermediate Griffiths regime for disorder strength $h\sim 3-4$, 
where the 
 long-range power-law entanglement of spins develops.  
The intermediate critical Griffiths regime  
is consistent with  some  earlier theoretical
studies using different probes\cite{agarwal2015, knap2015, vosk_theory2014, potter2015trans, serbyn2015}.
Distribution functions for spin-z  $P(S^z)$, entanglement entropy $P(S)$ and  fluctuations of the half system
magnetization $P(F)$ are  all distinctly  different from the ergodic phase or the MBL phase for finite-size systems, which also 
show  slow  evolution   with the system size $N$.
The basic physical picture revealed from our numerical studies is that 
the emergent conservation laws remain robust in the process of developing power-law long-range entanglements  
 in the critical Griffiths 
regime for  systems accessible by current numerical simulations.
The fate of the critical regime in the thermodynamic limit is unclear limited by the system sizes we study. However,
these results may provide new insights  for understanding the dynamic phase transition in such systems.

The En-DMRG algorithm we have  developed  is a new tool  for studying outstanding and challenging issues in
quantum statistical mechanics emerging in strongly interacting disorder systems. One of exciting directions
is to explore the nature of MBL phase in higher dimensions as the dimensionality always plays an essential
role in localization physics. Another direction is to explore the physics of the system with
the quantum order including the topological order in the MBL regime.
On the other hand, it is  still a challenge  to apply this method closer to the transition region
where the entanglement distribution is extremely broad, which we hope to  address in a separate work.
We also hope that our results  of characteristic  spin correlations and distribution functions  can stimulate   
experimental studies  along this direction.   

{\bf Acknowledgments} - DNS thanks David Huse  for stimulating discussions on DMRG for MBL states during her visit
at Princeton last winter, and we also thank him for the valuable comments on our paper.
 This work is supported by US National Science Foundation  Grants 
PREM DMR-1205734, DMR-1408560, and Princeton MRSEC Grant DMR-0819860 for travel support.


\begin{thebibliography}{75}
\expandafter\ifx\csname natexlab\endcsname\relax\def\natexlab#1{#1}\fi
\expandafter\ifx\csname bibnamefont\endcsname\relax
  \def\bibnamefont#1{#1}\fi
\expandafter\ifx\csname bibfnamefont\endcsname\relax
  \def\bibfnamefont#1{#1}\fi
\expandafter\ifx\csname citenamefont\endcsname\relax
  \def\citenamefont#1{#1}\fi
\expandafter\ifx\csname url\endcsname\relax
  \def\url#1{\texttt{#1}}\fi
\expandafter\ifx\csname urlprefix\endcsname\relax\def\urlprefix{URL }\fi
\providecommand{\bibinfo}[2]{#2}
\providecommand{\eprint}[2][]{\url{#2}}

\bibitem[{\citenamefont{{Basko} et~al.}(2006)\citenamefont{{Basko}, {Aleiner},
  and {Altshuler}}}]{basko2006}
\bibinfo{author}{\bibfnamefont{D.~M.} \bibnamefont{{Basko}}},
  \bibinfo{author}{\bibfnamefont{I.~L.} \bibnamefont{{Aleiner}}},
  \bibnamefont{and} \bibinfo{author}{\bibfnamefont{B.~L.}
  \bibnamefont{{Altshuler}}}, \bibinfo{journal}{Annals of Physics}
  \textbf{\bibinfo{volume}{321}}, \bibinfo{pages}{1126} (\bibinfo{year}{2006}).

\bibitem[{\citenamefont{{Fleishman} and {Anderson}}(1980)}]{fleishman1980}
\bibinfo{author}{\bibfnamefont{L.}~\bibnamefont{{Fleishman}}} \bibnamefont{and}
  \bibinfo{author}{\bibfnamefont{P.~W.} \bibnamefont{{Anderson}}},
  \bibinfo{journal}{\prb} \textbf{\bibinfo{volume}{21}}, \bibinfo{pages}{2366}
  (\bibinfo{year}{1980}).

\bibitem[{\citenamefont{{Altshuler} et~al.}(1997)\citenamefont{{Altshuler},
  {Gefen}, {Kamenev}, and {Levitov}}}]{altshuler1997}
\bibinfo{author}{\bibfnamefont{B.~L.} \bibnamefont{{Altshuler}}},
  \bibinfo{author}{\bibfnamefont{Y.}~\bibnamefont{{Gefen}}},
  \bibinfo{author}{\bibfnamefont{A.}~\bibnamefont{{Kamenev}}},
  \bibnamefont{and} \bibinfo{author}{\bibfnamefont{L.~S.}
  \bibnamefont{{Levitov}}}, \bibinfo{journal}{Phys. Rev. Lett.}
  \textbf{\bibinfo{volume}{78}}, \bibinfo{pages}{2803} (\bibinfo{year}{1997}).

\bibitem[{\citenamefont{{Jacquod} and {Shepelyansky}}(1997)}]{jacquod1997}
\bibinfo{author}{\bibfnamefont{P.}~\bibnamefont{{Jacquod}}} \bibnamefont{and}
  \bibinfo{author}{\bibfnamefont{D.~L.} \bibnamefont{{Shepelyansky}}},
  \bibinfo{journal}{Phys. Rev. Lett.} \textbf{\bibinfo{volume}{79}},
  \bibinfo{pages}{1837} (\bibinfo{year}{1997}).

\bibitem[{\citenamefont{{Georgeot} and {Shepelyansky}}(1998)}]{georgeot1998}
\bibinfo{author}{\bibfnamefont{B.}~\bibnamefont{{Georgeot}}} \bibnamefont{and}
  \bibinfo{author}{\bibfnamefont{D.~L.} \bibnamefont{{Shepelyansky}}},
  \bibinfo{journal}{Phys. Rev. Lett.} \textbf{\bibinfo{volume}{81}},
  \bibinfo{pages}{5129} (\bibinfo{year}{1998}).

\bibitem[{\citenamefont{{Gornyi} et~al.}(2005)\citenamefont{{Gornyi}, {Mirlin},
  and {Polyakov}}}]{gornyi2005}
\bibinfo{author}{\bibfnamefont{I.~V.} \bibnamefont{{Gornyi}}},
  \bibinfo{author}{\bibfnamefont{A.~D.} \bibnamefont{{Mirlin}}},
  \bibnamefont{and} \bibinfo{author}{\bibfnamefont{D.~G.}
  \bibnamefont{{Polyakov}}}, \bibinfo{journal}{Phys. Rev. Lett.}
  \textbf{\bibinfo{volume}{95}}, \bibinfo{eid}{206603} (\bibinfo{year}{2005}).

\bibitem[{\citenamefont{{Nandkishore} and {Huse}}(2015)}]{nandkishore2015}
\bibinfo{author}{\bibfnamefont{R.}~\bibnamefont{{Nandkishore}}}
  \bibnamefont{and} \bibinfo{author}{\bibfnamefont{D.~A.}
  \bibnamefont{{Huse}}}, \bibinfo{journal}{Annu. Rev. Cond. Matt. Phys.}
  \textbf{\bibinfo{volume}{6}}, \bibinfo{pages}{15} (\bibinfo{year}{2015}).

\bibitem[{\citenamefont{{Altman} and {Vosk}}(2015)}]{altman2015}
\bibinfo{author}{\bibfnamefont{E.}~\bibnamefont{{Altman}}} \bibnamefont{and}
  \bibinfo{author}{\bibfnamefont{R.}~\bibnamefont{{Vosk}}},
  \bibinfo{journal}{Annu. Rev. Cond. Matt. Phys.} \textbf{\bibinfo{volume}{6}},
  \bibinfo{pages}{383} (\bibinfo{year}{2015}).

\bibitem[{\citenamefont{{Huse} et~al.}(2014)\citenamefont{{Huse},
  {Nandkishore}, and {Oganesyan}}}]{huse2014}
\bibinfo{author}{\bibfnamefont{D.~A.} \bibnamefont{{Huse}}},
  \bibinfo{author}{\bibfnamefont{R.}~\bibnamefont{{Nandkishore}}},
  \bibnamefont{and}
  \bibinfo{author}{\bibfnamefont{V.}~\bibnamefont{{Oganesyan}}},
  \bibinfo{journal}{\prb} \textbf{\bibinfo{volume}{90}}, \bibinfo{eid}{174202}
  (\bibinfo{year}{2014}).

\bibitem[{\citenamefont{{Nandkishore} et~al.}(2014)\citenamefont{{Nandkishore},
  {Gopalakrishnan}, and {Huse}}}]{nandkishore2014}
\bibinfo{author}{\bibfnamefont{R.}~\bibnamefont{{Nandkishore}}},
  \bibinfo{author}{\bibfnamefont{S.}~\bibnamefont{{Gopalakrishnan}}},
  \bibnamefont{and} \bibinfo{author}{\bibfnamefont{D.~A.}
  \bibnamefont{{Huse}}}, \bibinfo{journal}{\prb} \textbf{\bibinfo{volume}{90}},
  \bibinfo{eid}{064203} (\bibinfo{year}{2014}).

\bibitem[{\citenamefont{{Oganesyan} and {Huse}}(2007)}]{oganesyan2007}
\bibinfo{author}{\bibfnamefont{V.}~\bibnamefont{{Oganesyan}}} \bibnamefont{and}
  \bibinfo{author}{\bibfnamefont{D.~A.} \bibnamefont{{Huse}}},
  \bibinfo{journal}{\prb} \textbf{\bibinfo{volume}{75}}, \bibinfo{eid}{155111}
  (\bibinfo{year}{2007}).

\bibitem[{\citenamefont{{Pal} and {Huse}}(2010)}]{pal2010}
\bibinfo{author}{\bibfnamefont{A.}~\bibnamefont{{Pal}}} \bibnamefont{and}
  \bibinfo{author}{\bibfnamefont{D.~A.} \bibnamefont{{Huse}}},
  \bibinfo{journal}{\prb} \textbf{\bibinfo{volume}{82}}, \bibinfo{eid}{174411}
  (\bibinfo{year}{2010}).

\bibitem[{\citenamefont{{{\v Z}nidari{\v c}} et~al.}(2008)\citenamefont{{{\v
  Z}nidari{\v c}}, {Prosen}, and {Prelov{\v s}ek}}}]{znidaric2008}
\bibinfo{author}{\bibfnamefont{M.}~\bibnamefont{{{\v Z}nidari{\v c}}}},
  \bibinfo{author}{\bibfnamefont{T.}~\bibnamefont{{Prosen}}}, \bibnamefont{and}
  \bibinfo{author}{\bibfnamefont{P.}~\bibnamefont{{Prelov{\v s}ek}}},
  \bibinfo{journal}{\prb} \textbf{\bibinfo{volume}{77}}, \bibinfo{eid}{064426}
  (\bibinfo{year}{2008}).

\bibitem[{\citenamefont{{Rigol} et~al.}(2008)\citenamefont{{Rigol}, {Dunjko},
  and {Olshanii}}}]{rigol2008}
\bibinfo{author}{\bibfnamefont{M.}~\bibnamefont{{Rigol}}},
  \bibinfo{author}{\bibfnamefont{V.}~\bibnamefont{{Dunjko}}}, \bibnamefont{and}
  \bibinfo{author}{\bibfnamefont{M.}~\bibnamefont{{Olshanii}}},
  \bibinfo{journal}{\nat} \textbf{\bibinfo{volume}{452}}, \bibinfo{pages}{854}
  (\bibinfo{year}{2008}).

\bibitem[{\citenamefont{{Serbyn} et~al.}(2014)\citenamefont{{Serbyn}, {Knap},
  {Gopalakrishnan}, {Papi{\'c}}, {Yao}, {Laumann}, {Abanin}, {Lukin}, and
  {Demler}}}]{serbyn2014}
\bibinfo{author}{\bibfnamefont{M.}~\bibnamefont{{Serbyn}}},
  \bibinfo{author}{\bibfnamefont{M.}~\bibnamefont{{Knap}}},
  \bibinfo{author}{\bibfnamefont{S.}~\bibnamefont{{Gopalakrishnan}}},
  \bibinfo{author}{\bibfnamefont{Z.}~\bibnamefont{{Papi{\'c}}}},
  \bibinfo{author}{\bibfnamefont{N.~Y.} \bibnamefont{{Yao}}},
  \bibinfo{author}{\bibfnamefont{C.~R.} \bibnamefont{{Laumann}}},
  \bibinfo{author}{\bibfnamefont{D.~A.} \bibnamefont{{Abanin}}},
  \bibinfo{author}{\bibfnamefont{M.~D.} \bibnamefont{{Lukin}}},
  \bibnamefont{and} \bibinfo{author}{\bibfnamefont{E.~A.}
  \bibnamefont{{Demler}}}, \bibinfo{journal}{Phys. Rev. Lett.}
  \textbf{\bibinfo{volume}{113}}, \bibinfo{eid}{147204} (\bibinfo{year}{2014}).

\bibitem[{\citenamefont{{Kwasigroch} and {Cooper}}(2014)}]{kwasigroch2014}
\bibinfo{author}{\bibfnamefont{M.~P.} \bibnamefont{{Kwasigroch}}}
  \bibnamefont{and} \bibinfo{author}{\bibfnamefont{N.~R.}
  \bibnamefont{{Cooper}}}, \bibinfo{journal}{\pra}
  \textbf{\bibinfo{volume}{90}}, \bibinfo{eid}{021605} (\bibinfo{year}{2014}).

\bibitem[{\citenamefont{{Yao} et~al.}(2014)\citenamefont{{Yao}, {Laumann},
  {Gopalakrishnan}, {Knap}, {M{\"u}ller}, {Demler}, and {Lukin}}}]{yao2014}
\bibinfo{author}{\bibfnamefont{N.~Y.} \bibnamefont{{Yao}}},
  \bibinfo{author}{\bibfnamefont{C.~R.} \bibnamefont{{Laumann}}},
  \bibinfo{author}{\bibfnamefont{S.}~\bibnamefont{{Gopalakrishnan}}},
  \bibinfo{author}{\bibfnamefont{M.}~\bibnamefont{{Knap}}},
  \bibinfo{author}{\bibfnamefont{M.}~\bibnamefont{{M{\"u}ller}}},
  \bibinfo{author}{\bibfnamefont{E.~A.} \bibnamefont{{Demler}}},
  \bibnamefont{and} \bibinfo{author}{\bibfnamefont{M.~D.}
  \bibnamefont{{Lukin}}}, \bibinfo{journal}{Phys. Rev. Lett.}
  \textbf{\bibinfo{volume}{113}}, \bibinfo{eid}{243002} (\bibinfo{year}{2014}).

\bibitem[{\citenamefont{{Vasseur} et~al.}(2015)\citenamefont{{Vasseur},
  {Parameswaran}, and {Moore}}}]{vasseur2015}
\bibinfo{author}{\bibfnamefont{R.}~\bibnamefont{{Vasseur}}},
  \bibinfo{author}{\bibfnamefont{S.~A.} \bibnamefont{{Parameswaran}}},
  \bibnamefont{and} \bibinfo{author}{\bibfnamefont{J.~E.}
  \bibnamefont{{Moore}}}, \bibinfo{journal}{\prb}
  \textbf{\bibinfo{volume}{91}}, \bibinfo{eid}{140202} (\bibinfo{year}{2015}).

\bibitem[{\citenamefont{{Vosk} et~al.}(2014)\citenamefont{{Vosk}, {Huse}, and
  {Altman}}}]{vosk_theory2014}
\bibinfo{author}{\bibfnamefont{R.}~\bibnamefont{{Vosk}}},
  \bibinfo{author}{\bibfnamefont{D.~A.} \bibnamefont{{Huse}}},
  \bibnamefont{and} \bibinfo{author}{\bibfnamefont{E.}~\bibnamefont{{Altman}}},
  \bibinfo{journal}{ArXiv e-prints}  (\bibinfo{year}{2014}),
  \eprint{1412.3117}.

\bibitem[{\citenamefont{{Serbyn} et~al.}(2013)\citenamefont{{Serbyn},
  {Papi{\'c}}, and {Abanin}}}]{serbyn2013}
\bibinfo{author}{\bibfnamefont{M.}~\bibnamefont{{Serbyn}}},
  \bibinfo{author}{\bibfnamefont{Z.}~\bibnamefont{{Papi{\'c}}}},
  \bibnamefont{and} \bibinfo{author}{\bibfnamefont{D.~A.}
  \bibnamefont{{Abanin}}}, \bibinfo{journal}{Phys. Rev. Lett.}
  \textbf{\bibinfo{volume}{111}}, \bibinfo{eid}{127201} (\bibinfo{year}{2013}).

\bibitem[{\citenamefont{{Ros} et~al.}(2015)\citenamefont{{Ros}, {M{\"u}ller},
  and {Scardicchio}}}]{ros2015}
\bibinfo{author}{\bibfnamefont{V.}~\bibnamefont{{Ros}}},
  \bibinfo{author}{\bibfnamefont{M.}~\bibnamefont{{M{\"u}ller}}},
  \bibnamefont{and}
  \bibinfo{author}{\bibfnamefont{A.}~\bibnamefont{{Scardicchio}}},
  \bibinfo{journal}{Nucl. Phys. B} \textbf{\bibinfo{volume}{891}},
  \bibinfo{pages}{420} (\bibinfo{year}{2015}).

\bibitem[{\citenamefont{{Chandran} et~al.}(2014)\citenamefont{{Chandran},
  {Khemani}, {Laumann}, and {Sondhi}}}]{chandran2014}
\bibinfo{author}{\bibfnamefont{A.}~\bibnamefont{{Chandran}}},
  \bibinfo{author}{\bibfnamefont{V.}~\bibnamefont{{Khemani}}},
  \bibinfo{author}{\bibfnamefont{C.~R.} \bibnamefont{{Laumann}}},
  \bibnamefont{and} \bibinfo{author}{\bibfnamefont{S.~L.}
  \bibnamefont{{Sondhi}}}, \bibinfo{journal}{\prb}
  \textbf{\bibinfo{volume}{89}}, \bibinfo{eid}{144201} (\bibinfo{year}{2014}).

\bibitem[{\citenamefont{{Grover}}(2014)}]{grover2014}
\bibinfo{author}{\bibfnamefont{T.}~\bibnamefont{{Grover}}},
  \bibinfo{journal}{ArXiv e-prints}  (\bibinfo{year}{2014}),
  \eprint{1405.1471}.

\bibitem[{\citenamefont{{Agarwal} et~al.}(2015)\citenamefont{{Agarwal},
  {Gopalakrishnan}, {Knap}, {M{\"u}ller}, and {Demler}}}]{agarwal2015}
\bibinfo{author}{\bibfnamefont{K.}~\bibnamefont{{Agarwal}}},
  \bibinfo{author}{\bibfnamefont{S.}~\bibnamefont{{Gopalakrishnan}}},
  \bibinfo{author}{\bibfnamefont{M.}~\bibnamefont{{Knap}}},
  \bibinfo{author}{\bibfnamefont{M.}~\bibnamefont{{M{\"u}ller}}},
  \bibnamefont{and} \bibinfo{author}{\bibfnamefont{E.}~\bibnamefont{{Demler}}},
  \bibinfo{journal}{Physical Review Letters} \textbf{\bibinfo{volume}{114}},
  \bibinfo{eid}{160401} (\bibinfo{year}{2015}), \eprint{1408.3413}.

\bibitem[{\citenamefont{{Gopalakrishnan}
  et~al.}(2015)\citenamefont{{Gopalakrishnan}, {Mueller}, {Khemani}, {Knap},
  {Demler}, and {Huse}}}]{knap2015}
\bibinfo{author}{\bibfnamefont{S.}~\bibnamefont{{Gopalakrishnan}}},
  \bibinfo{author}{\bibfnamefont{M.}~\bibnamefont{{Mueller}}},
  \bibinfo{author}{\bibfnamefont{V.}~\bibnamefont{{Khemani}}},
  \bibinfo{author}{\bibfnamefont{M.}~\bibnamefont{{Knap}}},
  \bibinfo{author}{\bibfnamefont{E.}~\bibnamefont{{Demler}}}, \bibnamefont{and}
  \bibinfo{author}{\bibfnamefont{D.~A.} \bibnamefont{{Huse}}},
  \bibinfo{journal}{ArXiv e-prints}  (\bibinfo{year}{2015}),
  \eprint{1502.07712}.

\bibitem[{\citenamefont{{Canovi} et~al.}(2011)\citenamefont{{Canovi},
  {Rossini}, {Fazio}, {Santoro}, and {Silva}}}]{canovi2011}
\bibinfo{author}{\bibfnamefont{E.}~\bibnamefont{{Canovi}}},
  \bibinfo{author}{\bibfnamefont{D.}~\bibnamefont{{Rossini}}},
  \bibinfo{author}{\bibfnamefont{R.}~\bibnamefont{{Fazio}}},
  \bibinfo{author}{\bibfnamefont{G.~E.} \bibnamefont{{Santoro}}},
  \bibnamefont{and} \bibinfo{author}{\bibfnamefont{A.}~\bibnamefont{{Silva}}},
  \bibinfo{journal}{\prb} \textbf{\bibinfo{volume}{83}}, \bibinfo{eid}{094431}
  (\bibinfo{year}{2011}).

\bibitem[{\citenamefont{{Cuevas} et~al.}(2012)\citenamefont{{Cuevas},
  {Feigel'Man}, {Ioffe}, and {Mezard}}}]{cuevas2012}
\bibinfo{author}{\bibfnamefont{E.}~\bibnamefont{{Cuevas}}},
  \bibinfo{author}{\bibfnamefont{M.}~\bibnamefont{{Feigel'Man}}},
  \bibinfo{author}{\bibfnamefont{L.}~\bibnamefont{{Ioffe}}}, \bibnamefont{and}
  \bibinfo{author}{\bibfnamefont{M.}~\bibnamefont{{Mezard}}},
  \bibinfo{journal}{Nat. Commun.} \textbf{\bibinfo{volume}{3}},
  \bibinfo{eid}{1128} (\bibinfo{year}{2012}).

\bibitem[{\citenamefont{{Bauer} and {Nayak}}(2013)}]{bauer2013}
\bibinfo{author}{\bibfnamefont{B.}~\bibnamefont{{Bauer}}} \bibnamefont{and}
  \bibinfo{author}{\bibfnamefont{C.}~\bibnamefont{{Nayak}}},
  \bibinfo{journal}{J. Stat. Mech. Theor. Exp.} \textbf{\bibinfo{volume}{9}},
  \bibinfo{eid}{09005} (\bibinfo{year}{2013}).

\bibitem[{\citenamefont{{Kj{\"a}ll} et~al.}(2014)\citenamefont{{Kj{\"a}ll},
  {Bardarson}, and {Pollmann}}}]{kjall2014}
\bibinfo{author}{\bibfnamefont{J.~A.} \bibnamefont{{Kj{\"a}ll}}},
  \bibinfo{author}{\bibfnamefont{J.~H.} \bibnamefont{{Bardarson}}},
  \bibnamefont{and}
  \bibinfo{author}{\bibfnamefont{F.}~\bibnamefont{{Pollmann}}},
  \bibinfo{journal}{Phys. Rev. Lett.} \textbf{\bibinfo{volume}{113}},
  \bibinfo{eid}{107204} (\bibinfo{year}{2014}).

\bibitem[{\citenamefont{{De Luca} and {Scardicchio}}(2013)}]{luca2013}
\bibinfo{author}{\bibfnamefont{A.}~\bibnamefont{{De Luca}}} \bibnamefont{and}
  \bibinfo{author}{\bibfnamefont{A.}~\bibnamefont{{Scardicchio}}},
  \bibinfo{journal}{EPL (Europhysics Letters)} \textbf{\bibinfo{volume}{101}},
  \bibinfo{pages}{37003} (\bibinfo{year}{2013}).

\bibitem[{\citenamefont{{Iyer} et~al.}(2013)\citenamefont{{Iyer}, {Oganesyan},
  {Refael}, and {Huse}}}]{iyer2013}
\bibinfo{author}{\bibfnamefont{S.}~\bibnamefont{{Iyer}}},
  \bibinfo{author}{\bibfnamefont{V.}~\bibnamefont{{Oganesyan}}},
  \bibinfo{author}{\bibfnamefont{G.}~\bibnamefont{{Refael}}}, \bibnamefont{and}
  \bibinfo{author}{\bibfnamefont{D.~A.} \bibnamefont{{Huse}}},
  \bibinfo{journal}{\prb} \textbf{\bibinfo{volume}{87}}, \bibinfo{eid}{134202}
  (\bibinfo{year}{2013}).

\bibitem[{\citenamefont{{Pekker} et~al.}(2014)\citenamefont{{Pekker}, {Refael},
  {Altman}, {Demler}, and {Oganesyan}}}]{pekker_hilbert2014}
\bibinfo{author}{\bibfnamefont{D.}~\bibnamefont{{Pekker}}},
  \bibinfo{author}{\bibfnamefont{G.}~\bibnamefont{{Refael}}},
  \bibinfo{author}{\bibfnamefont{E.}~\bibnamefont{{Altman}}},
  \bibinfo{author}{\bibfnamefont{E.}~\bibnamefont{{Demler}}}, \bibnamefont{and}
  \bibinfo{author}{\bibfnamefont{V.}~\bibnamefont{{Oganesyan}}},
  \bibinfo{journal}{Phys. Rev. X} \textbf{\bibinfo{volume}{4}},
  \bibinfo{eid}{011052} (\bibinfo{year}{2014}).

\bibitem[{\citenamefont{{Johri} et~al.}(2014)\citenamefont{{Johri},
  {Nandkishore}, and {Bhatt}}}]{johri2014}
\bibinfo{author}{\bibfnamefont{S.}~\bibnamefont{{Johri}}},
  \bibinfo{author}{\bibfnamefont{R.}~\bibnamefont{{Nandkishore}}},
  \bibnamefont{and} \bibinfo{author}{\bibfnamefont{R.~N.}
  \bibnamefont{{Bhatt}}}, \bibinfo{journal}{ArXiv e-prints}
  (\bibinfo{year}{2014}), \eprint{1405.5515}.

\bibitem[{\citenamefont{{Bardarson} et~al.}(2012)\citenamefont{{Bardarson},
  {Pollmann}, and {Moore}}}]{bardarson2012}
\bibinfo{author}{\bibfnamefont{J.~H.} \bibnamefont{{Bardarson}}},
  \bibinfo{author}{\bibfnamefont{F.}~\bibnamefont{{Pollmann}}},
  \bibnamefont{and} \bibinfo{author}{\bibfnamefont{J.~E.}
  \bibnamefont{{Moore}}}, \bibinfo{journal}{Phys. Rev. Lett.}
  \textbf{\bibinfo{volume}{109}}, \bibinfo{eid}{017202} (\bibinfo{year}{2012}).

\bibitem[{\citenamefont{{Andraschko} et~al.}(2014)\citenamefont{{Andraschko},
  {Enss}, and {Sirker}}}]{andraschko2014}
\bibinfo{author}{\bibfnamefont{F.}~\bibnamefont{{Andraschko}}},
  \bibinfo{author}{\bibfnamefont{T.}~\bibnamefont{{Enss}}}, \bibnamefont{and}
  \bibinfo{author}{\bibfnamefont{J.}~\bibnamefont{{Sirker}}},
  \bibinfo{journal}{Phys. Rev. Lett.} \textbf{\bibinfo{volume}{113}},
  \bibinfo{eid}{217201} (\bibinfo{year}{2014}).

\bibitem[{\citenamefont{{Laumann} et~al.}(2014)\citenamefont{{Laumann}, {Pal},
  and {Scardicchio}}}]{laumann2014}
\bibinfo{author}{\bibfnamefont{C.~R.} \bibnamefont{{Laumann}}},
  \bibinfo{author}{\bibfnamefont{A.}~\bibnamefont{{Pal}}}, \bibnamefont{and}
  \bibinfo{author}{\bibfnamefont{A.}~\bibnamefont{{Scardicchio}}},
  \bibinfo{journal}{Phys. Rev. Lett.} \textbf{\bibinfo{volume}{113}},
  \bibinfo{eid}{200405} (\bibinfo{year}{2014}).

\bibitem[{\citenamefont{{Hickey} et~al.}(2014)\citenamefont{{Hickey}, {Genway},
  and {Garrahan}}}]{hickey2014}
\bibinfo{author}{\bibfnamefont{J.~M.} \bibnamefont{{Hickey}}},
  \bibinfo{author}{\bibfnamefont{S.}~\bibnamefont{{Genway}}}, \bibnamefont{and}
  \bibinfo{author}{\bibfnamefont{J.~P.} \bibnamefont{{Garrahan}}},
  \bibinfo{journal}{ArXiv e-prints}  (\bibinfo{year}{2014}),
  \eprint{1405.5780}.

\bibitem[{\citenamefont{{Nanduri} et~al.}(2014)\citenamefont{{Nanduri}, {Kim},
  and {Huse}}}]{nanduri2014}
\bibinfo{author}{\bibfnamefont{A.}~\bibnamefont{{Nanduri}}},
  \bibinfo{author}{\bibfnamefont{H.}~\bibnamefont{{Kim}}}, \bibnamefont{and}
  \bibinfo{author}{\bibfnamefont{D.~A.} \bibnamefont{{Huse}}},
  \bibinfo{journal}{\prb} \textbf{\bibinfo{volume}{90}}, \bibinfo{eid}{064201}
  (\bibinfo{year}{2014}).

\bibitem[{\citenamefont{{Bar Lev} and {Reichman}}(2014)}]{barlev2014}
\bibinfo{author}{\bibfnamefont{Y.}~\bibnamefont{{Bar Lev}}} \bibnamefont{and}
  \bibinfo{author}{\bibfnamefont{D.~R.} \bibnamefont{{Reichman}}},
  \bibinfo{journal}{\prb} \textbf{\bibinfo{volume}{89}}, \bibinfo{eid}{220201}
  (\bibinfo{year}{2014}).

\bibitem[{\citenamefont{{Imbrie}}(2014)}]{imbrie2014}
\bibinfo{author}{\bibfnamefont{J.~Z.} \bibnamefont{{Imbrie}}},
  \bibinfo{journal}{ArXiv e-prints}  (\bibinfo{year}{2014}),
  \eprint{1403.7837}.

\bibitem[{\citenamefont{{Grover} and {Fisher}}(2014)}]{groverf2014}
\bibinfo{author}{\bibfnamefont{T.}~\bibnamefont{{Grover}}} \bibnamefont{and}
  \bibinfo{author}{\bibfnamefont{M.~P.~A.} \bibnamefont{{Fisher}}},
  \bibinfo{journal}{J. Stat. Mech. Theor. Exp.} \textbf{\bibinfo{volume}{10}},
  \bibinfo{eid}{10010} (\bibinfo{year}{2014}).

\bibitem[{\citenamefont{{Ponte} et~al.}(2015)\citenamefont{{Ponte},
  {Papi{\'c}}, {Huveneers}, and {Abanin}}}]{ponte2015}
\bibinfo{author}{\bibfnamefont{P.}~\bibnamefont{{Ponte}}},
  \bibinfo{author}{\bibfnamefont{Z.}~\bibnamefont{{Papi{\'c}}}},
  \bibinfo{author}{\bibfnamefont{F.}~\bibnamefont{{Huveneers}}},
  \bibnamefont{and} \bibinfo{author}{\bibfnamefont{D.~A.}
  \bibnamefont{{Abanin}}}, \bibinfo{journal}{Phys. Rev. Lett.}
  \textbf{\bibinfo{volume}{114}}, \bibinfo{eid}{140401} (\bibinfo{year}{2015}).

\bibitem[{\citenamefont{{Huang}}(2015)}]{huang2015}
\bibinfo{author}{\bibfnamefont{Y.}~\bibnamefont{{Huang}}},
  \bibinfo{journal}{ArXiv e-prints}  (\bibinfo{year}{2015}),
  \eprint{1507.01304}.

\bibitem[{\citenamefont{{You} et~al.}(2015)\citenamefont{{You}, {Qi}, and
  {Xu}}}]{you2015}
\bibinfo{author}{\bibfnamefont{Y.-Z.} \bibnamefont{{You}}},
  \bibinfo{author}{\bibfnamefont{X.-L.} \bibnamefont{{Qi}}}, \bibnamefont{and}
  \bibinfo{author}{\bibfnamefont{C.}~\bibnamefont{{Xu}}},
  \bibinfo{journal}{ArXiv e-prints}  (\bibinfo{year}{2015}),
  \eprint{1508.03635}.

\bibitem[{\citenamefont{{Serbyn} and {Moore}}(2015)}]{serbyn2015}
\bibinfo{author}{\bibfnamefont{M.}~\bibnamefont{{Serbyn}}} \bibnamefont{and}
  \bibinfo{author}{\bibfnamefont{J.~E.} \bibnamefont{{Moore}}},
  \bibinfo{journal}{ArXiv e-prints}  (\bibinfo{year}{2015}),
  \eprint{1508.07293}.

\bibitem[{\citenamefont{{Singh} et~al.}(2015)\citenamefont{{Singh},
  {Bardarson}, and {Pollmann}}}]{singh2015}
\bibinfo{author}{\bibfnamefont{R.}~\bibnamefont{{Singh}}},
  \bibinfo{author}{\bibfnamefont{J.~H.} \bibnamefont{{Bardarson}}},
  \bibnamefont{and}
  \bibinfo{author}{\bibfnamefont{F.}~\bibnamefont{{Pollmann}}},
  \bibinfo{journal}{ArXiv e-prints}  (\bibinfo{year}{2015}),
  \eprint{1508.05045}.

\bibitem[{\citenamefont{{Bar Lev} and {Reichman}}(2015)}]{barlev2015}
\bibinfo{author}{\bibfnamefont{Y.}~\bibnamefont{{Bar Lev}}} \bibnamefont{and}
  \bibinfo{author}{\bibfnamefont{D.~R.} \bibnamefont{{Reichman}}},
  \bibinfo{journal}{ArXiv e-prints}  (\bibinfo{year}{2015}),
  \eprint{1508.05391}.

\bibitem[{\citenamefont{{Deng} et~al.}(2015)\citenamefont{{Deng}, {Pixley},
  {Li}, and {Das Sarma}}}]{deng2015}
\bibinfo{author}{\bibfnamefont{D.-L.} \bibnamefont{{Deng}}},
  \bibinfo{author}{\bibfnamefont{J.~H.} \bibnamefont{{Pixley}}},
  \bibinfo{author}{\bibfnamefont{X.}~\bibnamefont{{Li}}}, \bibnamefont{and}
  \bibinfo{author}{\bibfnamefont{S.}~\bibnamefont{{Das Sarma}}},
  \bibinfo{journal}{ArXiv e-prints}  (\bibinfo{year}{2015}),
  \eprint{1508.01270}.

\bibitem[{\citenamefont{{Chen} et~al.}(2015)\citenamefont{{Chen}, {Yu}, {Cho},
  {Clark}, and {Fradkin}}}]{chen2015}
\bibinfo{author}{\bibfnamefont{X.}~\bibnamefont{{Chen}}},
  \bibinfo{author}{\bibfnamefont{X.}~\bibnamefont{{Yu}}},
  \bibinfo{author}{\bibfnamefont{G.~Y.} \bibnamefont{{Cho}}},
  \bibinfo{author}{\bibfnamefont{B.~K.} \bibnamefont{{Clark}}},
  \bibnamefont{and}
  \bibinfo{author}{\bibfnamefont{E.}~\bibnamefont{{Fradkin}}},
  \bibinfo{journal}{ArXiv e-prints}  (\bibinfo{year}{2015}),
  \eprint{1509.03890}.

\bibitem[{\citenamefont{Li et~al.}(2015)\citenamefont{Li, Ganeshan, Pixley, and
  Das~Sarma}}]{li2015}
\bibinfo{author}{\bibfnamefont{X.}~\bibnamefont{Li}},
  \bibinfo{author}{\bibfnamefont{S.}~\bibnamefont{Ganeshan}},
  \bibinfo{author}{\bibfnamefont{J.~H.} \bibnamefont{Pixley}},
  \bibnamefont{and}
  \bibinfo{author}{\bibfnamefont{S.}~\bibnamefont{Das~Sarma}},
  \bibinfo{journal}{Phys. Rev. Lett.} \textbf{\bibinfo{volume}{115}},
  \bibinfo{pages}{186601} (\bibinfo{year}{2015}).

\bibitem[{\citenamefont{{Potter} et~al.}(2015)\citenamefont{{Potter},
  {Vasseur}, and {Parameswaran}}}]{potter2015trans}
\bibinfo{author}{\bibfnamefont{A.~C.} \bibnamefont{{Potter}}},
  \bibinfo{author}{\bibfnamefont{R.}~\bibnamefont{{Vasseur}}},
  \bibnamefont{and} \bibinfo{author}{\bibfnamefont{S.~A.}
  \bibnamefont{{Parameswaran}}}, \bibinfo{journal}{ArXiv e-prints}
  (\bibinfo{year}{2015}), \eprint{1501.03501}.

\bibitem[{\citenamefont{{Deutsch}}(1991)}]{deutsch1991}
\bibinfo{author}{\bibfnamefont{J.~M.} \bibnamefont{{Deutsch}}},
  \bibinfo{journal}{\pra} \textbf{\bibinfo{volume}{43}}, \bibinfo{pages}{2046}
  (\bibinfo{year}{1991}).

\bibitem[{\citenamefont{{Srednicki}}(1994)}]{srednicki1994}
\bibinfo{author}{\bibfnamefont{M.}~\bibnamefont{{Srednicki}}},
  \bibinfo{journal}{\pre} \textbf{\bibinfo{volume}{50}}, \bibinfo{pages}{888}
  (\bibinfo{year}{1994}).

\bibitem[{\citenamefont{{Hosur} and {Qi}}(2015)}]{hosur2015}
\bibinfo{author}{\bibfnamefont{P.}~\bibnamefont{{Hosur}}} \bibnamefont{and}
  \bibinfo{author}{\bibfnamefont{X.-L.} \bibnamefont{{Qi}}},
  \bibinfo{journal}{ArXiv e-prints}  (\bibinfo{year}{2015}),
  \eprint{1507.04003}.

\bibitem[{\citenamefont{{Chandran}
  et~al.}(2015{\natexlab{a}})\citenamefont{{Chandran}, {Carrasquilla}, {Kim},
  {Abanin}, and {Vidal}}}]{chandran2015}
\bibinfo{author}{\bibfnamefont{A.}~\bibnamefont{{Chandran}}},
  \bibinfo{author}{\bibfnamefont{J.}~\bibnamefont{{Carrasquilla}}},
  \bibinfo{author}{\bibfnamefont{I.~H.} \bibnamefont{{Kim}}},
  \bibinfo{author}{\bibfnamefont{D.~A.} \bibnamefont{{Abanin}}},
  \bibnamefont{and} \bibinfo{author}{\bibfnamefont{G.}~\bibnamefont{{Vidal}}},
  \bibinfo{journal}{\prb} \textbf{\bibinfo{volume}{92}}, \bibinfo{eid}{024201}
  (\bibinfo{year}{2015}{\natexlab{a}}).

\bibitem[{\citenamefont{{Huse} et~al.}(2013)\citenamefont{{Huse},
  {Nandkishore}, {Oganesyan}, {Pal}, and {Sondhi}}}]{huse2013}
\bibinfo{author}{\bibfnamefont{D.~A.} \bibnamefont{{Huse}}},
  \bibinfo{author}{\bibfnamefont{R.}~\bibnamefont{{Nandkishore}}},
  \bibinfo{author}{\bibfnamefont{V.}~\bibnamefont{{Oganesyan}}},
  \bibinfo{author}{\bibfnamefont{A.}~\bibnamefont{{Pal}}}, \bibnamefont{and}
  \bibinfo{author}{\bibfnamefont{S.~L.} \bibnamefont{{Sondhi}}},
  \bibinfo{journal}{\prb} \textbf{\bibinfo{volume}{88}}, \bibinfo{eid}{014206}
  (\bibinfo{year}{2013}).

\bibitem[{\citenamefont{{Bahri} et~al.}(2013)\citenamefont{{Bahri}, {Vosk},
  {Altman}, and {Vishwanath}}}]{bahri2013}
\bibinfo{author}{\bibfnamefont{Y.}~\bibnamefont{{Bahri}}},
  \bibinfo{author}{\bibfnamefont{R.}~\bibnamefont{{Vosk}}},
  \bibinfo{author}{\bibfnamefont{E.}~\bibnamefont{{Altman}}}, \bibnamefont{and}
  \bibinfo{author}{\bibfnamefont{A.}~\bibnamefont{{Vishwanath}}},
  \bibinfo{journal}{ArXiv e-prints}  (\bibinfo{year}{2013}).

\bibitem[{\citenamefont{{Vosk} and {Altman}}(2014)}]{vosk2014}
\bibinfo{author}{\bibfnamefont{R.}~\bibnamefont{{Vosk}}} \bibnamefont{and}
  \bibinfo{author}{\bibfnamefont{E.}~\bibnamefont{{Altman}}},
  \bibinfo{journal}{Phys. Rev. Lett.} \textbf{\bibinfo{volume}{112}},
  \bibinfo{eid}{217204} (\bibinfo{year}{2014}).

\bibitem[{\citenamefont{{Potter} and {Vishwanath}}(2015)}]{potter2015}
\bibinfo{author}{\bibfnamefont{A.~C.} \bibnamefont{{Potter}}} \bibnamefont{and}
  \bibinfo{author}{\bibfnamefont{A.}~\bibnamefont{{Vishwanath}}},
  \bibinfo{journal}{ArXiv e-prints}  (\bibinfo{year}{2015}),
  \eprint{1506.00592}.

\bibitem[{\citenamefont{{Yao} et~al.}(2015)\citenamefont{{Yao}, {Laumann}, and
  {Vishwanath}}}]{yao2015}
\bibinfo{author}{\bibfnamefont{N.~Y.} \bibnamefont{{Yao}}},
  \bibinfo{author}{\bibfnamefont{C.~R.} \bibnamefont{{Laumann}}},
  \bibnamefont{and}
  \bibinfo{author}{\bibfnamefont{A.}~\bibnamefont{{Vishwanath}}},
  \bibinfo{journal}{ArXiv e-prints}  (\bibinfo{year}{2015}),
  \eprint{1508.06995}.

\bibitem[{\citenamefont{{Bordia} et~al.}(2015)\citenamefont{{Bordia},
  {L{\"u}schen}, {Hodgman}, {Schreiber}, {Bloch}, and
  {Schneider}}}]{bordia2015}
\bibinfo{author}{\bibfnamefont{P.}~\bibnamefont{{Bordia}}},
  \bibinfo{author}{\bibfnamefont{H.~P.} \bibnamefont{{L{\"u}schen}}},
  \bibinfo{author}{\bibfnamefont{S.~S.} \bibnamefont{{Hodgman}}},
  \bibinfo{author}{\bibfnamefont{M.}~\bibnamefont{{Schreiber}}},
  \bibinfo{author}{\bibfnamefont{I.}~\bibnamefont{{Bloch}}}, \bibnamefont{and}
  \bibinfo{author}{\bibfnamefont{U.}~\bibnamefont{{Schneider}}},
  \bibinfo{journal}{ArXiv e-prints}  (\bibinfo{year}{2015}),
  \eprint{1509.00478}.

\bibitem[{\citenamefont{{Pekker} and {Clark}}(2014{\natexlab{a}})}]{pekker2014}
\bibinfo{author}{\bibfnamefont{D.}~\bibnamefont{{Pekker}}} \bibnamefont{and}
  \bibinfo{author}{\bibfnamefont{B.~K.} \bibnamefont{{Clark}}},
  \bibinfo{journal}{ArXiv e-prints}  (\bibinfo{year}{2014}{\natexlab{a}}),
  \eprint{1410.2224}.

\bibitem[{\citenamefont{{Luitz} et~al.}(2015)\citenamefont{{Luitz},
  {Laflorencie}, and {Alet}}}]{luitz2015}
\bibinfo{author}{\bibfnamefont{D.~J.} \bibnamefont{{Luitz}}},
  \bibinfo{author}{\bibfnamefont{N.}~\bibnamefont{{Laflorencie}}},
  \bibnamefont{and} \bibinfo{author}{\bibfnamefont{F.}~\bibnamefont{{Alet}}},
  \bibinfo{journal}{\prb} \textbf{\bibinfo{volume}{91}}, \bibinfo{eid}{081103}
  (\bibinfo{year}{2015}).

\bibitem[{\citenamefont{{Goold} et~al.}(2015)\citenamefont{{Goold}, {Gogolin},
  {Clark}, {Eisert}, {Scardicchio}, and {Silva}}}]{goold2015}
\bibinfo{author}{\bibfnamefont{J.}~\bibnamefont{{Goold}}},
  \bibinfo{author}{\bibfnamefont{C.}~\bibnamefont{{Gogolin}}},
  \bibinfo{author}{\bibfnamefont{S.~R.} \bibnamefont{{Clark}}},
  \bibinfo{author}{\bibfnamefont{J.}~\bibnamefont{{Eisert}}},
  \bibinfo{author}{\bibfnamefont{A.}~\bibnamefont{{Scardicchio}}},
  \bibnamefont{and} \bibinfo{author}{\bibfnamefont{A.}~\bibnamefont{{Silva}}},
  \bibinfo{journal}{ArXiv e-prints}  (\bibinfo{year}{2015}),
  \eprint{1504.06872}.

\bibitem[{\citenamefont{{Devakul} and {Singh}}(2015)}]{devakul2015}
\bibinfo{author}{\bibfnamefont{T.}~\bibnamefont{{Devakul}}} \bibnamefont{and}
  \bibinfo{author}{\bibfnamefont{R.~R.~P.} \bibnamefont{{Singh}}},
  \bibinfo{journal}{ArXiv e-prints}  (\bibinfo{year}{2015}),
  \eprint{1508.04813}.

\bibitem[{\citenamefont{{Chandran}
  et~al.}(2015{\natexlab{b}})\citenamefont{{Chandran}, {Laumann}, and
  {Oganesyan}}}]{chandran2015finite}
\bibinfo{author}{\bibfnamefont{A.}~\bibnamefont{{Chandran}}},
  \bibinfo{author}{\bibfnamefont{C.~R.} \bibnamefont{{Laumann}}},
  \bibnamefont{and}
  \bibinfo{author}{\bibfnamefont{V.}~\bibnamefont{{Oganesyan}}},
  \bibinfo{journal}{ArXiv e-prints}  (\bibinfo{year}{2015}{\natexlab{b}}),
  \eprint{1509.04285}.

\bibitem[{\citenamefont{White}(1992)}]{white1992}
\bibinfo{author}{\bibfnamefont{S.~R.} \bibnamefont{White}},
  \bibinfo{journal}{Phys. Rev. Lett.} \textbf{\bibinfo{volume}{69}},
  \bibinfo{pages}{2863} (\bibinfo{year}{1992}).

\bibitem[{\citenamefont{{Pekker} and
  {Clark}}(2014{\natexlab{b}})}]{pekker2014_mps}
\bibinfo{author}{\bibfnamefont{D.}~\bibnamefont{{Pekker}}} \bibnamefont{and}
  \bibinfo{author}{\bibfnamefont{B.~K.} \bibnamefont{{Clark}}},
  \bibinfo{journal}{ArXiv e-prints}  (\bibinfo{year}{2014}{\natexlab{b}}),
  \eprint{1410.2224}.

\bibitem[{\citenamefont{{Friesdorf} et~al.}(2015)\citenamefont{{Friesdorf},
  {Werner}, {Brown}, {Scholz}, and {Eisert}}}]{friesdorf2015}
\bibinfo{author}{\bibfnamefont{M.}~\bibnamefont{{Friesdorf}}},
  \bibinfo{author}{\bibfnamefont{A.~H.} \bibnamefont{{Werner}}},
  \bibinfo{author}{\bibfnamefont{W.}~\bibnamefont{{Brown}}},
  \bibinfo{author}{\bibfnamefont{V.~B.} \bibnamefont{{Scholz}}},
  \bibnamefont{and} \bibinfo{author}{\bibfnamefont{J.}~\bibnamefont{{Eisert}}},
  \bibinfo{journal}{Phys. Rev. Lett.} \textbf{\bibinfo{volume}{114}},
  \bibinfo{eid}{170505} (\bibinfo{year}{2015}).

\bibitem[{\citenamefont{{Pollmann} et~al.}(2015)\citenamefont{{Pollmann},
  {Khemani}, {Cirac}, and {Sondhi}}}]{pollmann2015}
\bibinfo{author}{\bibfnamefont{F.}~\bibnamefont{{Pollmann}}},
  \bibinfo{author}{\bibfnamefont{V.}~\bibnamefont{{Khemani}}},
  \bibinfo{author}{\bibfnamefont{J.~I.} \bibnamefont{{Cirac}}},
  \bibnamefont{and} \bibinfo{author}{\bibfnamefont{S.~L.}
  \bibnamefont{{Sondhi}}}, \bibinfo{journal}{ArXiv e-prints}
  (\bibinfo{year}{2015}), \eprint{1506.07179}.

\bibitem[{\citenamefont{{Chandran}
  et~al.}(2015{\natexlab{c}})\citenamefont{{Chandran}, {Kim}, {Vidal}, and
  {Abanin}}}]{chandran2015_construct}
\bibinfo{author}{\bibfnamefont{A.}~\bibnamefont{{Chandran}}},
  \bibinfo{author}{\bibfnamefont{I.~H.} \bibnamefont{{Kim}}},
  \bibinfo{author}{\bibfnamefont{G.}~\bibnamefont{{Vidal}}}, \bibnamefont{and}
  \bibinfo{author}{\bibfnamefont{D.~A.} \bibnamefont{{Abanin}}},
  \bibinfo{journal}{\prb} \textbf{\bibinfo{volume}{91}}, \bibinfo{eid}{085425}
  (\bibinfo{year}{2015}{\natexlab{c}}).

\bibitem[{\citenamefont{{Khemani} et~al.}(2015)\citenamefont{{Khemani},
  {Pollmann}, and {Sondhi}}}]{khemani2015}
\bibinfo{author}{\bibfnamefont{V.}~\bibnamefont{{Khemani}}},
  \bibinfo{author}{\bibfnamefont{F.}~\bibnamefont{{Pollmann}}},
  \bibnamefont{and} \bibinfo{author}{\bibfnamefont{S.~L.}
  \bibnamefont{{Sondhi}}}, \bibinfo{journal}{ArXiv e-prints}
  (\bibinfo{year}{2015}), \eprint{1509.00483}.

\bibitem[{\citenamefont{{Yu} et~al.}(2015)\citenamefont{{Yu}, {Pekker}, and
  {Clark}}}]{yu2015}
\bibinfo{author}{\bibfnamefont{X.}~\bibnamefont{{Yu}}},
  \bibinfo{author}{\bibfnamefont{D.}~\bibnamefont{{Pekker}}}, \bibnamefont{and}
  \bibinfo{author}{\bibfnamefont{B.~K.} \bibnamefont{{Clark}}},
  \bibinfo{journal}{ArXiv e-prints}  (\bibinfo{year}{2015}),
  \eprint{1509.01244}.

\bibitem[{\citenamefont{{Jiang} et~al.}(2012)\citenamefont{{Jiang}, {Wang}, and
  {Balents}}}]{jiang2012}
\bibinfo{author}{\bibfnamefont{H.-C.} \bibnamefont{{Jiang}}},
  \bibinfo{author}{\bibfnamefont{Z.}~\bibnamefont{{Wang}}}, \bibnamefont{and}
  \bibinfo{author}{\bibfnamefont{L.}~\bibnamefont{{Balents}}},
  \bibinfo{journal}{Nature Physics} \textbf{\bibinfo{volume}{8}},
  \bibinfo{pages}{902} (\bibinfo{year}{2012}), \eprint{1205.4289}.

\bibitem[{\citenamefont{Motrunich et~al.}(2000)\citenamefont{Motrunich, Mau,
  Huse, and Fisher}}]{motrunich2000}
\bibinfo{author}{\bibfnamefont{O.}~\bibnamefont{Motrunich}},
  \bibinfo{author}{\bibfnamefont{S.-C.} \bibnamefont{Mau}},
  \bibinfo{author}{\bibfnamefont{D.~A.} \bibnamefont{Huse}}, \bibnamefont{and}
  \bibinfo{author}{\bibfnamefont{D.~S.} \bibnamefont{Fisher}},
  \bibinfo{journal}{Phys. Rev. B} \textbf{\bibinfo{volume}{61}},
  \bibinfo{pages}{1160} (\bibinfo{year}{2000}).

\end{thebibliography}

\end{document}